%% file: Look.tex
\documentclass[sigconf, screen]{acmart}
\usepackage{enumitem}
\usepackage{multicol}
\usepackage{soul}
\usepackage{hyperref}       
\usepackage{url}            
\usepackage{booktabs}       
\usepackage{array}          
\usepackage{multirow}       
\usepackage{xcolor}         
\usepackage{colortbl} 
\usepackage{cuted,tcolorbox}
\usepackage{xspace}
\usepackage{arydshln}
\usepackage{subcaption}

\definecolor{oursorange}{RGB}{255,244,230}
\definecolor{bestred}{RGB}{252,228,236}
\definecolor{betterblue}{RGB}{232,242,254}

\newcommand{\bestcell}[1]{\cellcolor{bestred}\textbf{#1}}
\newcommand{\bettercell}[1]{\cellcolor{betterblue}#1}
\newcommand{\imp}[1]{\cellcolor{oursorange}#1}
\newcommand{\implast}[1]{\multicolumn{1}{>{\columncolor{oursorange}}c}{#1}}

\newcommand{\tool}[1]{\textsc{#1}\xspace}
\newcommand{\sysname}{\tool{FixATE}}
\newcommand{\backbone}{\tool{Backbone}}

\AtBeginDocument{%
  }

\setcopyright{acmlicensed}
\copyrightyear{2018}
\acmYear{2018}
\acmDOI{XXXXXXX.XXXXXXX}

\acmConference[Conference acronym 'XX]{}{June 03--05,
  2018}{Woodstock, NY}
\acmISBN{978-1-4503-XXXX-X/18/06}

\begin{document}

\title{Through Their Eyes: Fixation-aligned Tuning \\ for Personalized User Emulation}
\renewcommand{\shorttitle}{Through Their Eyes: Fixation-aligned Tuning for Personalized User Emulation}

\input{Main/authors}

\renewcommand{\shortauthors}{Huang et al.}

\begin{abstract}
Large language model (LLM) agents are increasingly deployed as scalable user simulators for recommender system evaluation.
Yet existing simulators perceive recommendations through text or structured metadata rather than the visual interfaces real users browse—a critical gap, since attention over recommendation layouts is both visually driven and highly personalized.
We investigate whether aligning a vision-language model's (VLM's) visual attention with user-specific gaze patterns can improve simulation fidelity.
Analysis of a real-world eye-tracking dataset collected in a carousel-based recommendation setting reveals that users exhibit stable individual gaze patterns strongly predictive of click behavior.
Building on this finding, we propose Fixation-Aligned Tuning for user Emulation (FixATE).
Our approach first probes the VLM’s internal visual attention via interpretability operators to obtain a slot-level relevance distribution comparable with human fixation, and then learns personalized soft prompts to steer the model’s attention toward each user’s characteristic fixation pattern.
Experiments across three interpretability-based probing operators and two architecturally distinct VLM backbones demonstrate consistent improvements in both attention alignment and click prediction accuracy.
These results suggest that making the model ``see like the user'' is a viable path toward simulators that more faithfully reproduce how users perceive and act in recommendation interfaces.
\end{abstract}



\maketitle

\input{Main/Introduction}
\input{Main/Motivation}

\input{Main/Methodology}
\input{Main/Experiments}
\input{Main/Related_works}

\input{Main/Conclusion}


\bibliographystyle{ACM-Reference-Format}
\bibliography{LOOK}

\clearpage
\appendix
\input{Main/Supplementary}


\end{document}

%% file: Main/authors.tex
\author{Lingfeng Huang}
\orcid{0000-0003-0861-5301}
\authornote{Equal contribution.}
\affiliation{%
  \institution{Singapore University of Technology and Design}
  \country{Singapore}
}
\email{lingfeng_huang@mymail.sutd.edu.sg}

\author{Huizhong Guo}
\orcid{0009-0004-0011-8612}
\authornotemark[1]
\affiliation{%
  \institution{Zhejiang University}
  \city{Hangzhou}
  \country{China}
}
\email{huiz_g@zju.edu.cn}

\author{Tianjun Wei}
\authornote{Corresponding author.}
\orcid{0000-0001-7311-7101}
\affiliation{%
  \institution{Nanyang Technological University}
  \country{Singapore}
}
\email{tjwei2-c@my.cityu.edu.hk}

\author{Yingpeng Du}
\orcid{0000-0001-9881-7171}
\affiliation{%
  \institution{Nanyang Technological University}
  \country{Singapore}
}
\email{dyp1993@pku.edu.cn}

\author{Zhu Sun}
\orcid{0000-0002-3350-7022}
\affiliation{%
  \institution{Singapore University of Technology and Design}
  \country{Singapore}
}
\email{zhu_sun@sutd.edu.sg}

%% file: Main/Introduction.tex
\section{Introduction}

Large language models (LLMs) have demonstrated remarkable capability as autonomous agents~\cite{zhangMultiMindEnhancing2025,liuEyeSherlock2025}.
A typical LLM agent operates through a \textbf{\textit{perception--memory--action}} loop: it perceives environmental stimuli, stores and retrieves relevant knowledge from memory, and takes actions that alter the environment \cite{wangAgentWorkflow2025}.
A growing body of work deploys LLM-based agents as efficient substitutes for human participants in evaluation-intensive scenarios, including review assessment \cite{zhengJudgingLLMasajudge2023}, web navigation \cite{zhouWebArenaRealistic2023}, and GUI operating \cite{zhouMAIUITechnical2025}.
Among these, the \textbf{recommender system} scenario stands out as a representative and widely explored setting, where \textbf{LLM-based user simulators} are increasingly recognized as a cost-effective and scalable alternative to human evaluation \cite{zhangGenerativeAgents2024,zhangLLMpoweredUser2025,zhuHowReliable2024,bougieSimUSERSimulating2025,liuDiagnosticGuidedDynamic2026, weiMirroringUsers2025, chenLassoLarge2025}.
These simulators construct detailed user profiles from historical interactions that record both preferences and affective states \cite{zhangGenerativeAgents2024,wangUserBehavior2025,zhangSurveyMemory2025}.
Equipped with such profiling and memory mechanisms, they have shown promising fidelity in reproducing user behaviors \cite{zhangGenerativeAgents2024,pengSurveyLLMpowered2025}.

\begin{figure}[t]
    \centering
    \includegraphics[width=0.85\linewidth]{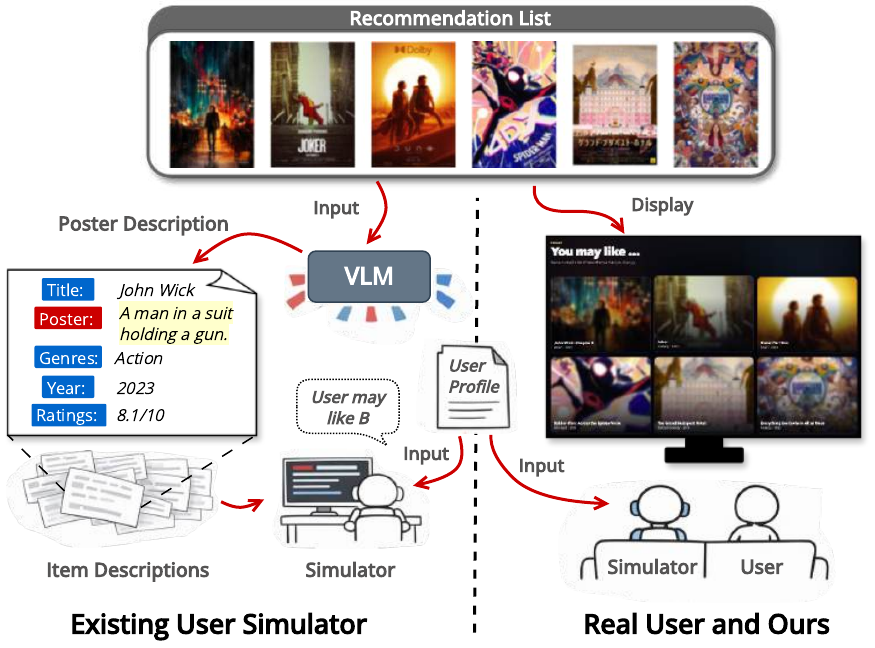}
    \caption{The difference of the perception of recommendations between real users and LLM-based user simulators.}
    \Description{
        This figure shows the difference of the perception of recommendations between real users and LLM-based user simulators.
        The right panel shows the real users' attention distribution when browsing the movie poster carousel.
        The left panel shows the LLM-based user simulator's perception: movie posters are converted into text descriptions and fed into the LLM user simulators.
    }
    \label{fig:task_difference}
    \vspace{-0.6cm}
\end{figure}

Despite their advances in memory and behavioral fidelity, a significant asymmetry remains in \textbf{how these simulators perceive the recommendation environment}.
Most existing simulators consume textual item descriptions or structured metadata as their sole sensory input.
Although recent work has begun integrating vision-language models (VLMs) into recommendation pipelines \cite{yuVisionLanguageRecommendationAttribute2019,jiGeneratingNegativeSamples2025,yadaImprovingVisual2025,wuAligningLarge2025,wu2026efficient} and user simulators \cite{bougieSimUSERSimulating2025}, these approaches still treat visual content as raw material to be converted into textual or embedding-level representations \textbf{rather than as the perceptual interface}.
In contrast, as shown in Figure \ref{fig:task_difference}, real users interact with rich visual presentations---movie posters arranged in a horizontally scrollable carousel, product thumbnails laid out on a grid shelf \cite{deleon-martinezRecGazeFirst2025,deleon-martinezRidingCarousel2026}.
Controlled eye-tracking experiments \cite{pfeifferEyeTrackingBasedClassification2020, loonPredictingProduct2022} confirm that visual presentation plays a central role in shaping initial attention allocation and purchase decisions.
\textbf{The gap between item-level visual processing and human-like visual interaction} indicates that existing VLM-equipped simulators may systematically miss the attentional signals that drive real user behavior.

To close this remaining gap, a parallel line of research seeks to align VLM visual attention with human gaze patterns.
Existing approaches include editing the input or prompt \cite{wanContrastiveRegion2024,yuAttentionPrompting2024}, reshaping internal attention distributions of vision encoders~\cite{paniGazeVLMBridging2025}, and introducing gaze-based training objectives to align model attention with human gaze heatmaps \cite{paniGazeVLMBridging2025,yanVoilaAAligning2024}.
These methods have demonstrated gains in medical imaging \cite{zhuGuidingMedical2025}, visual question answering \cite{yanVoilaAAligning2024} and understanding \cite{paniGazeVLMBridging2025}.
However, \textbf{all existing visual attention alignment methods for VLMs operate at a task-level or universal level}. They learn a single, population-averaged attention pattern, disregarding the well-documented finding that different individuals viewing the same image attend to markedly different regions depending on their preferences and goals \cite{strohmLearningUser2024,gonzalezGenderawareSaliency2025}.
To our knowledge, the entire line of gaze-guided VLM alignment research has yet to enter the recommendation setting, where personalization is not a secondary concern but the defining challenge \cite{deleon-martinezRecGazeFirst2025,yadaImprovingVisual2025}.

To verify this hypothesis, we conduct an empirical analysis on a real-world eye-tracking dataset \cite{deleon-martinezRecGazeFirst2025} collected in a visual recommendation scenario.
Our analysis reveals that individual users exhibit distinct and stable gaze patterns when browsing item carousels \cite{deleon-martinezRecGazeFirst2025,strohmLearningUser2024}, and that these per-user attention patterns are strongly correlated with subsequent click behavior.
This observation is consistent with findings in adjacent domains.
Eye-tracking studies on search engine result pages have documented substantial individual differences in viewing strategies and attention allocation \cite{lewandowskiFactorsInfluencing2021}, retail gaze research has shown that fixation dwell time reliably predicts product preference at the individual level \cite{loonPredictingProduct2022}, and user-embedding studies have demonstrated that personal gaze patterns generalize across unseen images and carry transferable preference information \cite{strohmLearningUser2024}.
More broadly, recent work on gaze-based reward modeling has confirmed that individual-level gaze signals provide a meaningful supervision signal for aligning model outputs with human intent \cite{lopez-cardonaSeeingEye2024}.
Together, these findings provide direct motivation for a personalized approach to visual attention alignment.

Building on this insight, we propose \textbf{Fixation-Aligned Tuning for user Emulation (FixATE\footnote{Our implementation is publicly available at \url{https://github.com/hijack-lf/FixATE}.})}, a personalized VLM-based user simulator whose visual perception is aligned with each user’s personalized attention pattern.
Our approach addresses two key questions: \textbf{how to measure} the alignment between human fixation and the model's internal visual attention, and \textbf{how to achieve} personalized fixation alignment for each user.
For the first, we leverage VLM interpretability operators to extract a slot-level visual relevance distribution that is directly comparable with human fixation.
For the second, we propose \textbf{Personalized Fixation-aligned Tuning}, which learns a shared prompt basis together with personalized coefficients to construct personalized soft prompts that steer the VLM's attention toward each user's characteristic fixation pattern.

\textbf{To our knowledge, our work is the first to apply personalized gaze-guided alignment to VLMs in the recommendation setting.}
We evaluate the framework across two architecturally distinct VLM backbones~\cite{baiQwen3VLTechnical2025,wangInternVL35Advancing2025} and three interpretability-based probing operators~\cite{abnarQuantifyingAttention2020,shenGLIMPSEHolistic2025,achtibatAttnLRPAttentionaware2024} to mitigate model-specific confounds.
Our experiments demonstrate that personalized fixation alignment effectively shifts the VLM's visual attention distribution toward each user's characteristic gaze pattern, and that this perceptual alignment contributes to improved consistency between the simulator's predicted decisions and the user's actual behavior, yielding gains on both attention-level and behavior-level alignment.
These results suggest that making the model ``see like the user'' is a viable and effective pathway to making it ``act like the user'', shedding light on the critical role of perceptual alignment in achieving human-like user simulation.

%% file: Main/Motivation.tex
\section{How Should a User Simulator Perceive Recommendations?}
\label{sec:motivation}

User simulators built on LLMs are increasingly deployed as cost-effective substitutes for human participants across various domains \cite{zhengJudgingLLMasajudge2023,zhouWebArenaRealistic2023,zhouMAIUITechnical2025}.
Among these, the recommender system setting represents a particularly active application area where user simulators have attracted growing attention.
By emulating realistic browsing and decision-making behaviors, they provide a scalable and cost-effective alternative to human evaluation for testing recommendation policies~\cite{zhangGenerativeAgents2024,zhangLLMpoweredUser2025}.
However, existing work has largely focused on memory and profiling mechanisms while paying little attention to \emph{how} the recommendation environment is presented to the simulator.
In this section, we present two empirical analyses on the RecGaze dataset~\cite{deleon-martinezRecGazeFirst2025}, which becomes a promising motivation for our proposed framework.

RecGaze contains real-world human interaction data collected in a movie recommendation scenario, where participants browse a carousel interface arranged in a $3 \times 5$ grid. 
In each session, participants browse the interface and then perform an action, such as clicking on a movie or scrolling. 
The dataset records both eye-tracking fixation data and cursor trajectories throughout each session. We retain all sessions in which the user's final action is a click on one of the displayed movies.

\subsection{Revealing the Perceptual Gap in Text-based Interface Processing}
\label{sec:perceptual_gap}
Existing VLM-equipped user simulators typically process visual recommendation content by converting interface images into structured textual descriptions~\cite{bougieSimUSERSimulating2025}.
While this conversion preserves semantic item-level information, it discards the spatial layout and inter-item visual context that real users rely on when making decisions---signals that eye-tracking research has shown to be primary drivers of attention allocation and choice behavior in both retail~\cite{pfeifferEyeTrackingBasedClassification2020, loonPredictingProduct2022} and carousel-based recommendation settings~\cite{deleon-martinezRidingCarousel2026}.

To quantify this perceptual gap, we design a controlled experiment to compare the click position distributions produced by a VLM (\textit{Qwen3-VL-4B-Instruct}~\cite{baiQwen3VLTechnical2025}) under three input modalities against the ground-truth human distribution observed in RecGaze.
\begin{itemize}[leftmargin=1.6em, itemsep=2pt, topsep=2pt, parsep=0pt]
    \item In the \textbf{Image} condition, the VLM receives the original screenshot of the recommendation interface.
    \item In the \textbf{Text w/o Pos.} condition, the interface is converted into a textual list of movie titles and poster descriptions without any positional information (e.g., column and row indices).
    \item In the \textbf{Text w/ Pos.} condition, the same textual list is augmented with explicit row and column indices for each item.
\end{itemize}

\begin{figure}[t!]
    \centering
    \begin{subfigure}[t]{\linewidth}
        \centering
        \includegraphics[width=0.85\linewidth]{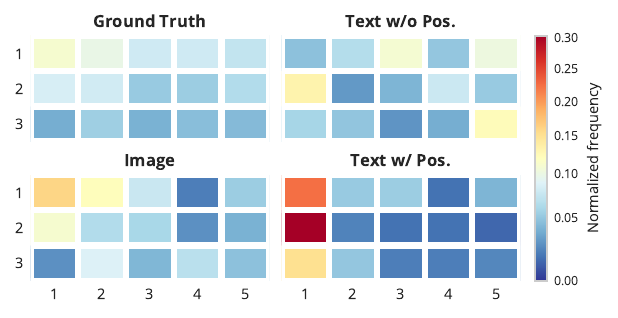}
        \captionsetup{skip=0.7pt}
        \caption{Heatmaps of the normalized click frequency of each slot.}
        \label{fig:position_heatmaps}
    \end{subfigure}
    \vspace{0.9em}
    \begin{subfigure}[t]{\linewidth}
        \centering
        \includegraphics[width=0.85\linewidth]{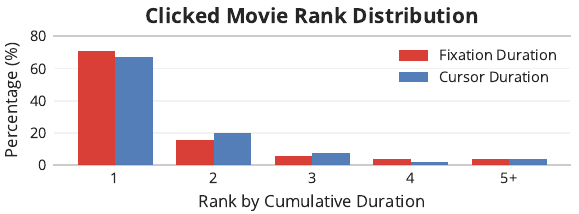}
        \caption{Distribution of the clicked movie's rank by cumulative fixation and cursor dwell time.}
        \label{fig:rank_distribution}
    \end{subfigure}
    \captionsetup{skip=0.7pt}
    \caption{Empirical analyses on the RecGaze dataset.}
    \Description{
        Two subfigures. (a) Four heatmaps showing click position distributions on a 3x5 grid for Ground Truth, Text w/o Pos., Image, and Text w/ Pos. conditions. The Image condition best matches the ground truth F-shaped pattern. (b) A grouped bar chart showing that the clicked movie ranks first in cumulative fixation duration 71\% of the time and first in cursor duration 67\% of the time.
    }
    \label{fig:motivation}
    \vspace{-1em}
\end{figure}

Figure~\ref{fig:motivation}(a) presents heatmaps of the normalized click frequency for each condition.
The ground-truth distribution exhibits a characteristic \textit{F-shaped pattern}, with elevated click rates in the upper-left region of the grid, consistent with well-documented primacy and position bias effects in carousel interfaces~\cite{deleon-martinezRidingCarousel2026}.
The \textbf{Image} condition most closely reproduces this pattern, indicating that when the VLM perceives the actual visual layout, it captures the spatial regularities that shape human click behavior.
In contrast, the \textbf{Text w/o Pos.} condition produces an irregular distribution skewed toward the last listed position, consistent with the ``lost-in-the-middle'' recency effect observed in long-context LLMs~\cite{liuLostMiddle2024}.
Adding explicit positional tags in the \textbf{Text w/ Pos.} condition overcorrects the problem: the model develops an exaggerated bias toward the first column, producing a distribution that is even more sharply concentrated than the ground truth, indicating an over-amplified positional bias.

These results demonstrate that \textbf{text-based representations of visual interfaces fail to faithfully convey the spatial information that shapes real user behavior}.
Providing the visual interface directly as input allows the VLM to implicitly capture layout-dependent behavioral patterns that textual workarounds either miss or distort in practice.

\subsection{Equipping Gaze and Cursor Signals as Behavioral Indicators}
\label{sec:gaze_behavior}
Having established that the visual interface should serve as the primary perceptual input, we next examine the relationship between pre-decisional attention signals and user behavior.
Prior work has established that fixation dwell time reliably predicts purchase decisions~\cite{loonPredictingProduct2022}, that individual viewing strategies on search result pages reflect stable personal habits~\cite{lewandowskiFactorsInfluencing2021}, that cursor position is strongly correlated with gaze position during web browsing~\cite{huangUserSee2012,chenWhatCan2001}, and that individual-level gaze signals can serve as effective supervision for aligning model outputs with human intent~\cite{lopez-cardonaSeeingEye2024}.

To directly examine this relationship in a visual recommendation setting, we analyze the same 575 click sessions from RecGaze.
For each session, we compute the cumulative fixation duration and cursor dwell duration on each of the 15 visible movie slots, and rank the movies by these cumulative durations.
Figure~\ref{fig:motivation}(b) presents the distribution of the clicked movie's rank, where  in more than 90\% of sessions, the clicked movie is the one with the Top-2 longest cumulative fixation and cursor dwell time.
This strong correspondence between pre-decisional attention allocation and the user's ultimate choice suggests that \textbf{if a user simulator's internal visual attention can be made to resemble that of a real user, its predicted actions are likely to become more behaviorally faithful}.
This insight forms the central motivation for our proposed framework: rather than treating visual attention alignment as a downstream evaluation metric, we elevate it to a first-class training objective.

%% file: Main/Methodology.tex
\section{Methodology}

Motivated by the empirical findings in the previous section, we propose to improve the perceptual fidelity of VLM-based user simulators by aligning their visual attention with human fixation behavior at the individual user level.
To realize this goal, we must address two key questions:
\textbf{(Q1)} How can we measure the alignment between human fixation and the model's internal visual attention?
\textbf{(Q2)} How can we achieve personalized fixation alignment that captures each user's characteristic viewing behavior?
In the following, we first formalize the user simulation task and establish notation (Sec.~\ref{sec:formulation}), then present our answers to Q1 (Sec.~\ref{sec:probing}) and Q2 (Sec.~\ref{sec:alignment}).

\subsection{Problem Formulation}
\label{sec:formulation}

We consider a visual recommendation scenario in which a user $u \in \mathcal{U}$ browses multiple recommendation sessions $t \in \mathcal{T}$.
In each session, the user is presented with an interface image that arranges $N$ candidate items in any layout.
Each area-of-interest (AOI) on the interface (denoted as a \emph{slot} for simplicity) corresponds to one item; we write $\mathcal{S} = \{1,\dots,N\}$ for the set of slot indices.

\noindent\textbf{User Simulation as Autoregressive Generation.}
Given a frozen vision-language model $\mathcal{M}$, the user simulator receives a multimodal context comprising the interface image $\mathbf{C}_{img}$, a textual user profile $\mathbf{C}_{prof}$, and a task instruction $\mathbf{C}_{inst}$, and generates a behavioral response through autoregressive decoding:
\begin{equation}
\small
p(y_{1:T} \mid \mathbf{C}_{img}, \mathbf{C}_{prof}, \mathbf{C}_{inst}) = \prod\nolimits_{i=1}^{T} p_{\mathcal{M}}(y_i \mid y_{<i}, \mathbf{C}_{img}, \mathbf{C}_{prof}, \mathbf{C}_{inst}),
\label{eq:autoregressive}
\end{equation}
where $y_{1:T}$ denotes the output token sequence. For example, if the target behavior is to make a click decision, the output is a single answer token $y_{u,t} \in \mathcal{S}$ that indicates the selected slot.

\noindent\textbf{Fixation Data.}
During browsing, an eye-tracking device records the user's gaze behavior as pixel-level fixation data. By aggregating dwell time within each slot's area of interest (AOI), we obtain a slot-level dwell-time vector $\mathbf{d}_{u,t} \in \mathbb{R}_{\ge 0}^{N}$, where each entry $d_{u,t}^{(n)}$ represents the total fixation dwell time on slot $n$.

\begin{figure*}[ht!]
    \centering
    \includegraphics[width=0.8\linewidth]{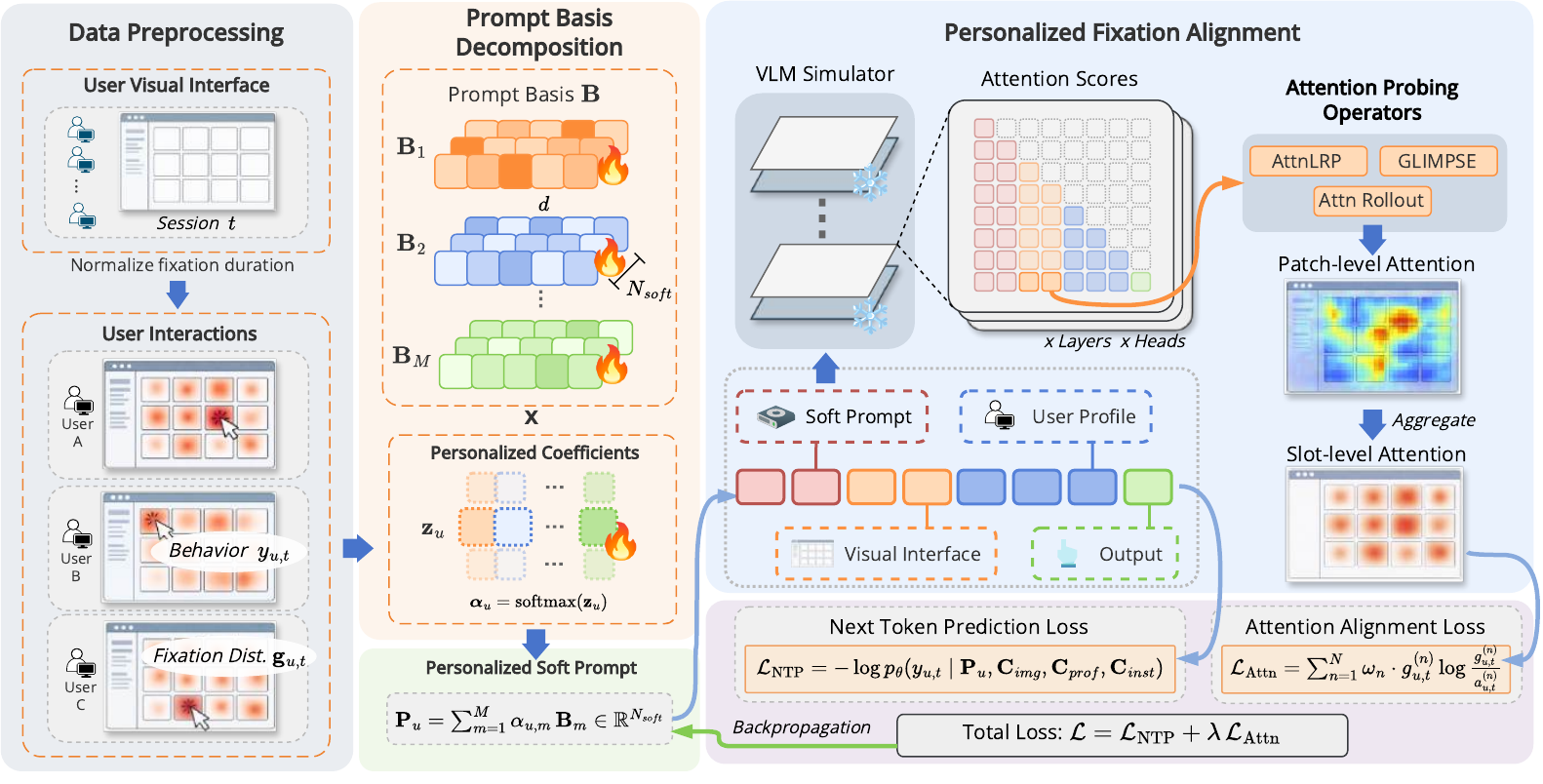}
    \captionsetup{skip=1.2pt}
    \caption{Overview of the Fixation-aligned Tuning for User Emulation (\sysname) framework.}
    \label{fig:framework}
\end{figure*}

\subsection{Probing Human-Aligned Visual Fixation}
\label{sec:probing}

To measure the alignment between human fixation and model attention, we need a model-side counterpart to the human dwell-time vector---a slot-level relevance distribution that reflects the model's internal visual focus over the recommendation interface. Recent VLM interpretability methods~\cite{shenGLIMPSEHolistic2025,achtibatAttnLRPAttentionaware2024,abnarQuantifyingAttention2020} offer a practical foundation for obtaining such a signal. Building on these methods, we define a general probing formulation and then describe the aggregation from token-level relevance to a slot-level distribution for downstream alignment analysis.

Formally, let the model $\mathcal{M}$ process a multimodal context $\mathbf{C}$ as input and produce an output sequence $\mathbf{y}$ through autoregressive decoding.
We define a probing function
\begin{equation}
\small
\label{eq:probing_function}
\mathbf{R} = f(\mathcal{M},\, \mathbf{C},\, \mathbf{y}) \in \mathbb{R}^{N_{\text{out}} \times N_{\text{cxt}}},
\end{equation}
where $\mathbf{R}$ is a relevance score matrix that quantifies the contribution of each input token to each output token.
Since our interest lies in \emph{visual} perception, we restrict $\mathbf{R}$ to the visual input tokens indexed by $\mathcal{V} \subset \{1,\dots,N_{\text{cxt}}\}$ and the output tokens corresponding to the generated behavior $\mathbf{y}$.
This yields the visual relevance matrix
$\mathbf{R}_{\mathcal{V}} \in \mathbb{R}^{N_{\text{out}} \times |\mathcal{V}|}$.
When the output contains multiple tokens, we average across the output dimension to obtain a single visual relevance vector $\mathbf{r}_{\text{vis}} \in \mathbb{R}^{|\mathcal{V}|}$, capturing the model's overall visual focus.

Our framework is agnostic to the choice of probing operator.
We instantiate $f$ with three representative interpretability methods: \textbf{Attention Rollout}~\cite{abnarQuantifyingAttention2020}, which approximates cross-layer influence propagation using attention weights alone; \textbf{GLIMPSE}~\cite{shenGLIMPSEHolistic2025}, which combines attention weights with task-specific gradients for output-conditioned relevance estimation; and \textbf{AttnLRP}~\cite{achtibatAttnLRPAttentionaware2024}, which propagates relevance through the attention structure via layer-wise relevance propagation.
Detailed descriptions of these operators are provided in Sec.~\ref{sec:experimental_setting}, together with implementation details.

Analogous to the pixel-level to slot-level aggregation used for human fixation data (Sec.~\ref{sec:formulation}), we map the patch-level visual relevance $\mathbf{r}_{\text{vis}}$ to a slot-level relevance vector $\mathbf{a}_{u,t} \in \mathbb{R}_{\ge 0}^{N}$ by assigning each image patch to its slot based on the layout geometry and summing the relevance values within each slot.

\subsection{Personalized Fixation Alignment}
\label{sec:alignment}

Given the slot-level relevance signal derived from the probing step, we now address how to steer the model toward each user's characteristic fixation pattern.
Our approach comprises two components: a \emph{personalized prompt basis decomposition} that constructs user-specific soft prompts (Sec.~\ref{sec:prompt_basis}), and a \emph{fixation-aligned tuning} objective that optimizes these prompts for both fixation alignment and behavioral prediction (Sec.~\ref{sec:fixation_tuning}).
Figure~\ref{fig:framework} illustrates the overall architecture of the framework.

\subsubsection{Personalized Prompt Basis Decomposition}
\label{sec:prompt_basis}


Through the probing procedure, we obtain slot-level fixation distributions from both humans and the VLM over the same interface layout for comparison. Prior literature on visual attention~\cite{xuPersonalizedSaliency2019,strohmLearningUser2024} suggests that these distributions contain both \emph{common} viewing tendencies shared across users (e.g., center bias, top-left primacy) and \emph{personalized} attentional patterns that reflect user-specific browsing habits and preferences. To accommodate this decomposition, we adopt a factorized soft prompt design~\cite{mengBlackBoxTestTime2025} that separates shared attentional prototypes from learnable user-specific composition weights.

\noindent\textbf{Prompt Basis.}
We define a shared fixation prompt token basis $\mathbf{B} \in \mathbb{R}^{M \times N_{soft} \times d}$, where $M$ is the number of basis prompts, $N_{soft}$ is the token sequence length per basis, and $d$ is the VLM input embedding dimension.
Each slice $\mathbf{B}_{m} \in \mathbb{R}^{N_{soft} \times d}$ contains $N_{soft}$ learnable soft token embeddings that capture a prototypical attentional mode shared across users, for personalized prompt construction.

\noindent\textbf{Personalized Coefficients.}
Each user $u$ is associated with a learnable logit vector $\mathbf{z}_u \in \mathbb{R}^{M}$, from which personalized mixing coefficients are obtained via softmax normalization:
\begin{equation}
\small
\boldsymbol{\alpha}_u = \mathrm{softmax}(\mathbf{z}_u) \in \Delta^{M-1}.
\label{eq:user-mixing-coefficients}
\end{equation}
The personalized soft prompt for user $u$ is then computed as a convex combination of the basis prompts:
\begin{equation}
\small
\mathbf{P}_u = \sum\nolimits_{m=1}^{M} \alpha_{u,m}\,\mathbf{B}_{m} \in \mathbb{R}^{N_{soft} \times d}.
\label{eq:soft-prompt}
\end{equation}
This design learns a dictionary of shared prompt basis $\mathbf{B}$ and, for each user, a personalized coefficient $\boldsymbol{\alpha}_u$ on the probability simplex, thereby generating user-specific soft prompts within a shared low-dimensional prompt subspace for efficient personalization.

The resulting personalized prompt $\mathbf{P}_u$ is prepended to the original input sequence to steer the VLM's attention toward user-specific fixation patterns.
All backbone parameters of $\mathcal{M}$ remain frozen; only the basis $\mathbf{B}$ and the user logit vectors $\{\mathbf{z}_u\}_{u \in \mathcal{U}}$ are trainable.

\subsubsection{Fixation-Aligned Tuning}
\label{sec:fixation_tuning}

With the personalized prompt mechanism in place, we define the training objective that jointly aligns the model's visual relevance with human fixation and maintains accurate behavioral prediction.

\noindent\textbf{Fixation Distribution Preparation.}
Given the human dwell-time vector $\mathbf{d}_{u,t}$, we first normalize the human fixation into a proper distribution via $\ell_1$ normalization to eliminate cross-session scale differences in total dwell time:
\begin{equation}
\small
\mathbf{g}_{u,t} = \mathbf{d}_{u,t} \cdot (\|\mathbf{d}_{u,t}\|_1 + \epsilon)^{-1},
\label{eq:g_ut}
\end{equation}
where $\epsilon > 0$ ensures numerical stability.

\noindent\textbf{Attention Alignment Loss.}
We employ a power-weighted Kullback–Leibler (KL) divergence~\cite{yangWeightedKLDivergenceDocument2024} to align the slot-level model relevance $\mathbf{a}_{u,t}$ with the human fixation distribution $\mathbf{g}_{u,t}$. In addition, an importance weight $\omega_n$ is added to control the emphasis on the most-attended slots. The loss function is defined as:
\begin{equation}
\small
\mathcal{L}_{\text{Attn}} = \sum_{n=1}^{N} \omega_n \cdot g_{u,t}^{(n)} \log \frac{g_{u,t}^{(n)}}{a_{u,t}^{(n)}},\:\:
\omega_n = \frac{(g_{u,t}^{(n)} + \epsilon)^{\gamma}}{\sum_{n'=1}^{N} (g_{u,t}^{(n')} + \epsilon)^{\gamma}},
\label{eq:attn-weight}
\end{equation}
and $\gamma \ge 0$ is a power exponent that controls how strongly the loss emphasizes high-gaze slots. When $\gamma = 0$, this formulation reduces to the standard KL divergence; as $\gamma$ increases, the loss places greater emphasis on aligning the most-attended slots.

\noindent\textbf{Next-Token Prediction Loss.}
In addition to fixation-level supervision, we apply the standard next-token prediction (NTP) loss under teacher forcing:
\begin{equation}
\small
\mathcal{L}_{\text{NTP}} = -\log p_\theta(y_{u,t} \mid \mathbf{P}_u, \mathbf{C}_{img}, \mathbf{C}_{prof}, \mathbf{C}_{inst}).
\label{eq:ntp-loss}
\end{equation}

\noindent\textbf{Total Loss.}
The final objective combines both losses:
\begin{equation}
\small
\mathcal{L} = \mathcal{L}_{\text{NTP}} + \lambda \, \mathcal{L}_{\text{Attn}},
\label{eq:total-loss}
\end{equation}
where $\lambda$ controls the relative importance of the fixation alignment loss.
Through this joint objective, the personalized soft prompts are optimized to simultaneously steer the model's internal visual attention toward human-like fixation patterns and preserve accurate click prediction, all without updating the backbone parameters.

%% file: Main/Experiments.tex
\section{Experiments}
\label{sec:experiments}
\subsection{Experimental Setting}
\label{sec:experimental_setting}
\textbf{Dataset.}
We conduct experiments on \textbf{RecGaze\footnote{\textbf{Ethics Statement.} This study is conducted using only the publicly released, de-identified version of RecGaze, with formal authorization from the dataset providers.}}~\cite{deleon-martinezRecGazeFirst2025}, a public eye-tracking dataset collected in a carousel-based visual recommendation interface. 
In RecGaze, participants browse movie recommendation pages arranged in a $3 \times 5$ grid and select items by clicking. 
The dataset records per-slot fixation dwell times, cursor trajectories, and final clicks for each session. 
As shown in Figure~\ref{fig:motivation}(b), the slot with the longest fixation dwell time coincides with the ultimately clicked slot more frequently than the cursor-based alternative. 
Moreover, fixation provides a more direct signal of visual attention, whereas cursor position can be displaced by scrolling, hovering, or other incidental motor actions. 
We therefore adopt fixation as the primary supervision signal and retain 575 interactions from 83 users with both fixation and click data for the main experiments.

We adopt a leave-one-out per-user strategy~\cite{maCIRPCrossItem2024}: for each user with at least two interactions, one sample is held out for testing and the rest are used for training. 
Users with fewer than two interactions are used only for training. For the remaining users, we hold out one interaction per user for testing and use the rest for training. 
This results in 76 test interactions and 499 training interactions. 

\textbf{VLM Backbones.}
We evaluate \sysname\ on two architecturally distinct VLM backbones, \textit{InternVL3.5-4B-Instruct}~\cite{wangInternVL35Advancing2025} and \textit{Qwen3-VL-4B-Instruct}~\cite{baiQwen3VLTechnical2025}, to examine its generality. During training, backbone parameters are frozen, and only the \sysname\ prompt parameters are optimized. Both models use \texttt{bfloat16} precision. Each forward pass receives a unified multimodal prompt with three components: \textbf{User Interface}, \textbf{User Profile}, and \textbf{Instruction}. To ensure reproducibility, the full prompt template is shown in Figure~\ref{fig:multimodal_prompt}.
\input{Tables/prompt}

\input{Tables/main_results}
\textbf{Probing Operators.}
\label{sec:exp_setup}
The framework \sysname\ is agnostic to the choice of probing operator $\Phi_m$ for relevance extraction. We instantiate it with three representative interpretability methods.
\begin{itemize}[leftmargin=1.6em, itemsep=2pt, topsep=2pt, parsep=0pt]
    \item \textbf{GLIMPSE}~\cite{shenGLIMPSEHolistic2025} is a method that combines attention weights with task-specific gradients to estimate relevance. It uses adaptive cross-layer aggregation to refine relevance propagation. This design preserves end-to-end gradient flow, allowing the attention alignment loss to directly optimize the soft prompts.
    \item \textbf{AttnLRP}~\cite{achtibatAttnLRPAttentionaware2024} is a relevance propagation method that attributes relevance through the transformer attention structure. Compared with \textit{GLIMPSE}, it uses a more stable depth-weighted aggregation scheme rather than iterative matrix products. This formulation is numerically more stable while still maintaining a differentiable path to the prompt parameters.
    \item \textbf{Attention Rollout}~\cite{abnarQuantifyingAttention2020} provides a post-hoc approximation of token importance by rolling out attention across transformer layers while accounting for residual connections. It serves as a lightweight baseline for tracing information propagation from input tokens to higher-layer representations.
\end{itemize}

\textbf{Evaluation Metrics.}
We employ two groups of metrics to evaluate alignment quality from complementary perspectives. 
\textbf{1) Attention alignment metrics} evaluate how well the probing-derived slot-level relevance distribution $\mathbf{a}_{u,t}$ matches user attention from two perspectives. First, \emph{global distribution alignment} measures the similarity between $\mathbf{a}_{u,t}$ and the user gaze distribution $\mathbf{g}_{u,t}$ using \emph{KL Divergence} (KL $\downarrow$), \emph{Jensen--Shannon Divergence} (JS $\downarrow$), and \emph{Cosine Similarity}. Second, \emph{top-slot alignment} assesses whether the model correctly identifies the user's most salient slots, using \emph{Clicked-Slot Hit@$k$} (CSH), which indicates whether the clicked slot is ranked within the model's top-$k$ slots according to the slot-level relevance scores in Sec.~\ref{sec:probing}, and \emph{Top-Gaze Overlap@$k$} (TGO), which measures the overlap between the model's and the user's top-$k$ slots.
\textbf{2) Prediction-level metrics} evaluate the final click prediction quality, including \emph{Accuracy}, which indicates whether the predicted answer matches the ground-truth click; \emph{LogLoss} ($\downarrow$), which measures the negative log-likelihood of the ground-truth click; and \emph{AUC}, which evaluates whether the clicked slot is ranked above non-clicked slots, regardless of the absolute score values. 

For all metrics, $\downarrow$ indicates that lower values are better, while metrics without this symbol are better when larger.
All reported results are averaged over 5 runs for each experimental case to reduce randomness. This protocol is adopted to improve the stability and reliability of the reported results in our experiments.

\textbf{Hyperparameter settings.}
Across all probing operators, we set the soft prompt length to $N_{soft}=16$, basis size to $M=8$, and use a power-weighted KL divergence with $\gamma=2$ for attention alignment. 
Training uses a batch size of 4 with gradient accumulation over 2 steps (effective batch size 8) for up to 30 epochs. Hyperparameters are selected via 5-fold cross-validation based on both attention-level and behavior-level metrics.
Detailed operator-specific settings are available in our open-source code.

\subsection{Experimental Results}
\label{sec:main_results}

Table~\ref{tab:main_results} presents the evaluation results of \sysname and \backbone (without personalized soft prompts) across three probing operators and two VLM backbones, under a unified experimental setting, covering attention alignment and prediction-level metrics. The main findings are summarized as follows. 

\textit{\textbf{First, the results consistently show that FixATE outperforms Backbone across all experiments, indicating that our personalized prompting strategy effectively improves human-attention alignment.}} As highlighted by the bold entries in Table 1, FixATE achieves consistently better performance than Backbone on nearly all metrics, including attention alignment metrics such as Cosine, JS, and KL, as well as prediction metrics such as Accuracy, LogLoss, and AUC. These results suggest that the learned user-specific prompt does not simply perturb relevance scores, but systematically steers the VLM toward attention patterns more consistent with real user behavior in recommendation scenarios. Overall, these findings provide strong and direct evidence that personalized attention alignment can be effectively induced in VLM-based user simulation, supporting the core premise of our method.

\textbf{\textit{Second, larger gains in top-slot attention alignment are associated with larger gains in downstream prediction accuracy.}} Although different probing operators are not directly comparable in absolute values, their relative gains over \backbone still provide a meaningful indicator of personalization-induced alignment improvement. From this perspective, \textit{settings with the largest CSH@1 gains, highlighted in red in Table~\ref{tab:main_results}, also tend to yield the largest Accuracy gains.} For example, under \textit{Qwen3-VL} with \textit{AttnLRP}, CSH@1 increases by 233.33\%, while Accuracy doubles, indicating that better identification of the behavior-relevant slot improves prediction reliability. Since CSH@1 measures whether the clicked slot is ranked first, this suggests that aligning the model with the user's most salient attention target is especially important for prediction. In other words, better top-slot alignment is not merely an auxiliary attention metric, but a key factor in accurate behavior modeling.

\textit{\textbf{Finally, the gains of \sysname remain stable across two VLM backbones and three probing operators, demonstrating strong generality and robustness. }} Across all probing operators and VLM backbones, introducing the learned personalized prompt consistently improves both alignment and prediction quality. This broad effectiveness suggests that our method does not rely on a specific architecture or probing mechanism, but instead functions as a model-agnostic attention steering framework. From a practical perspective, this universality is desirable because \sysname can be readily integrated into different VLM-based user simulation pipelines without modifying the backbone or probing strategy. In summary, these results show that \sysname is not only effective but also broadly applicable, highlighting its value as a general framework for improving human-aligned VLM-based user simulation.

\input{Tables/ablation}

\subsection{Ablation Study}
\label{sec:ablation}
Table~\ref{tab:ablation} reports the ablation results of \sysname with \textit{AttnLRP} on two VLM backbones. In addition to the full model, we consider several reduced variants to isolate the effect of each component. Specifically, \textit{Rand.~SP} replaces the learned soft prompt with a randomly initialized one; $\boldsymbol{z_{u}} = \mathbf{1}$ removes personalization; \textit{w/o $\omega_n$} removes the importance weights (Equation~\ref{eq:attn-weight}); and the \textit{w/o $\mathcal{L}_{\text{Attn}}/\mathcal{L}_{\text{NTP}}$} variants remove the corresponding loss. 
In the table, the best value within each model block is marked in light red, while values that outperform \backbone are highlighted in light blue.

\textit{\textbf{The ablation results show that each component contributes to overall performance.}} 
The full model achieves the strongest or near-strongest results on most metrics across both backbones, indicating that the gains of \sysname arise from the interaction of personalized prompting, weighted attention alignment, and joint optimization rather than any single component. 
In contrast, \textit{Rand.~SP} generally fails to improve over \backbone and often degrades performance, suggesting that the benefit comes not from adding prompt parameters alone, but from learning a structured and user-adaptive prompt. 
Similarly, removing user-specific coefficients $\boldsymbol{\alpha}_u$ consistently weakens performance, confirming that \textit{personalization is essential rather than optional for faithful user emulation}.

A particularly clear pattern is that removing the attention alignment objective leads to substantial degradation, whereas removing the NTP objective causes a moderate decline. This suggests that \textit{\textbf{the primary gains come from aligning the model’s relevance distribution with human attention}}. Without $\mathcal{L}_{\text{Attn}}$, both alignment and prediction metrics drop noticeably, indicating that attention alignment provides the main training signal for learning human-aligned slot salience. By contrast, the variant without $\mathcal{L}_{\text{NTP}}$ often remains competitive and even attains the highest values on some top-slot metrics such as TGO@1, suggesting that the attention objective alone provides a strong foundation. However, the full model still achieves better overall Accuracy, LogLoss, and AUC in all cases, suggesting that \textit{$\mathcal{L}_{\text{NTP}}$ plays a complementary role by refining aligned attention toward the final clicked slot and translating better grounding into stronger behavior prediction}.
\begin{figure}[t]
    \centering
    \includegraphics[width=\linewidth]{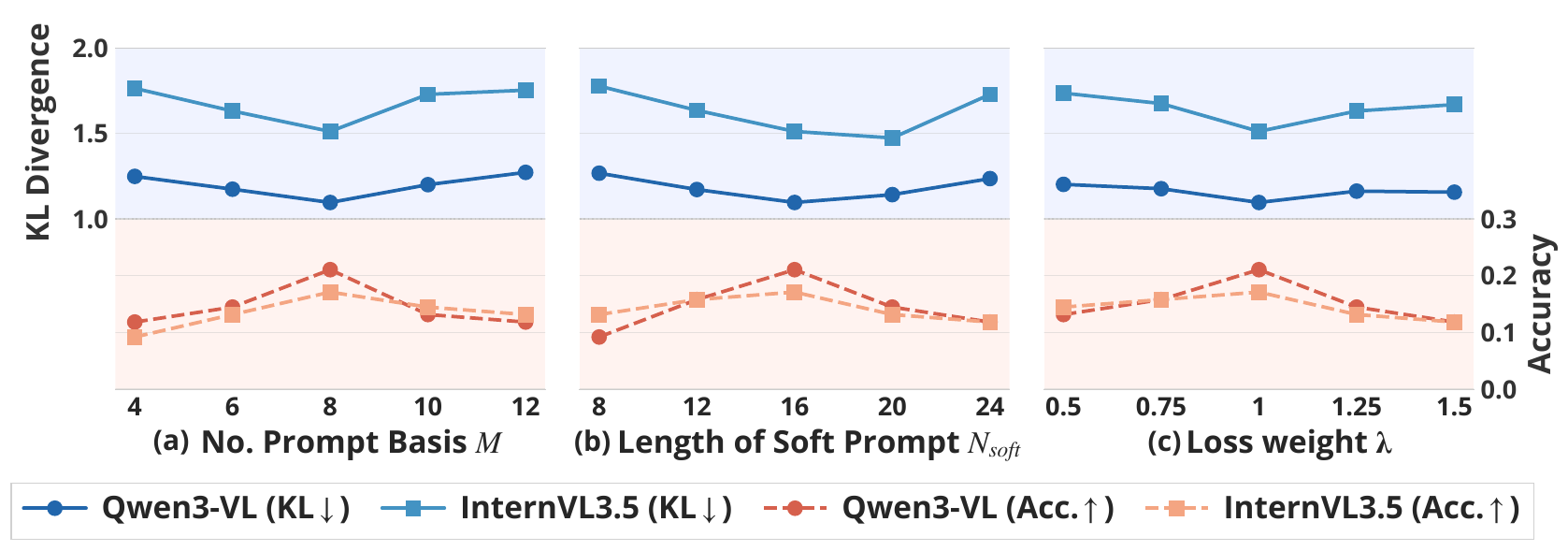}
    \captionsetup{skip=1.3pt}
    \caption{Sensitivity analysis of key hyperparameters on two backbones with \textit{AttnLRP}. Each curve reports a representative metric as the hyperparameter varies.}
    \label{fig:sensitivity}
    \vspace{-1em}
\end{figure}

\subsection{Hyperparameter Sensitivity}
Figure~\ref{fig:sensitivity} presents a sensitivity analysis of three key hyperparameters:
the number of basis prompts $M$, the number of tokens per basis prompt
$N_\text{soft}$, and the loss trade-off weight $\lambda$.
We report KL divergence and Accuracy under \textit{AttnLRP} on both VLM backbones.

Both capacity-related hyperparameters exhibit a consistent non-monotonic trend across
backbones: increasing prompt capacity initially improves performance, but overly large
values yield diminishing returns or degradation.
This suggests that the prompt bank requires sufficient expressiveness to capture
user-specific attention patterns, while excessive capacity introduces redundancy and
destabilizes optimization.
Performance peaks at $M = 8$ and $N_\text{soft} = 16$, where both models
achieve the lowest KL divergence and the highest or near-highest Accuracy.
Notably, \textit{InternVL3.5} attains slightly lower KL divergence at $N_\text{soft} = 20$,
yet its Accuracy declines---indicating that distributional alignment and behavior
prediction do not always move in lockstep, and that the chosen configuration provides
the best overall balance.

The $\lambda$ panel reveals a similar alignment--prediction trade-off.
Performance peaks around $\lambda = 1.0$ on both backbones: too small a value
underweights the fixation signal and leaves slot-level relevance insufficiently shaped,
while too large a value over-emphasizes alignment at the expense of click prediction.
Based on these results, we adopt $M = 8$, $N_\text{soft} = 16$, and
$\lambda = 1.0$ in all main experiments.

\subsection{Case Study}
\begin{figure}[t]
    \centering
    \begin{subfigure}[t]{0.32\linewidth}
        \centering
        \includegraphics[width=\linewidth]{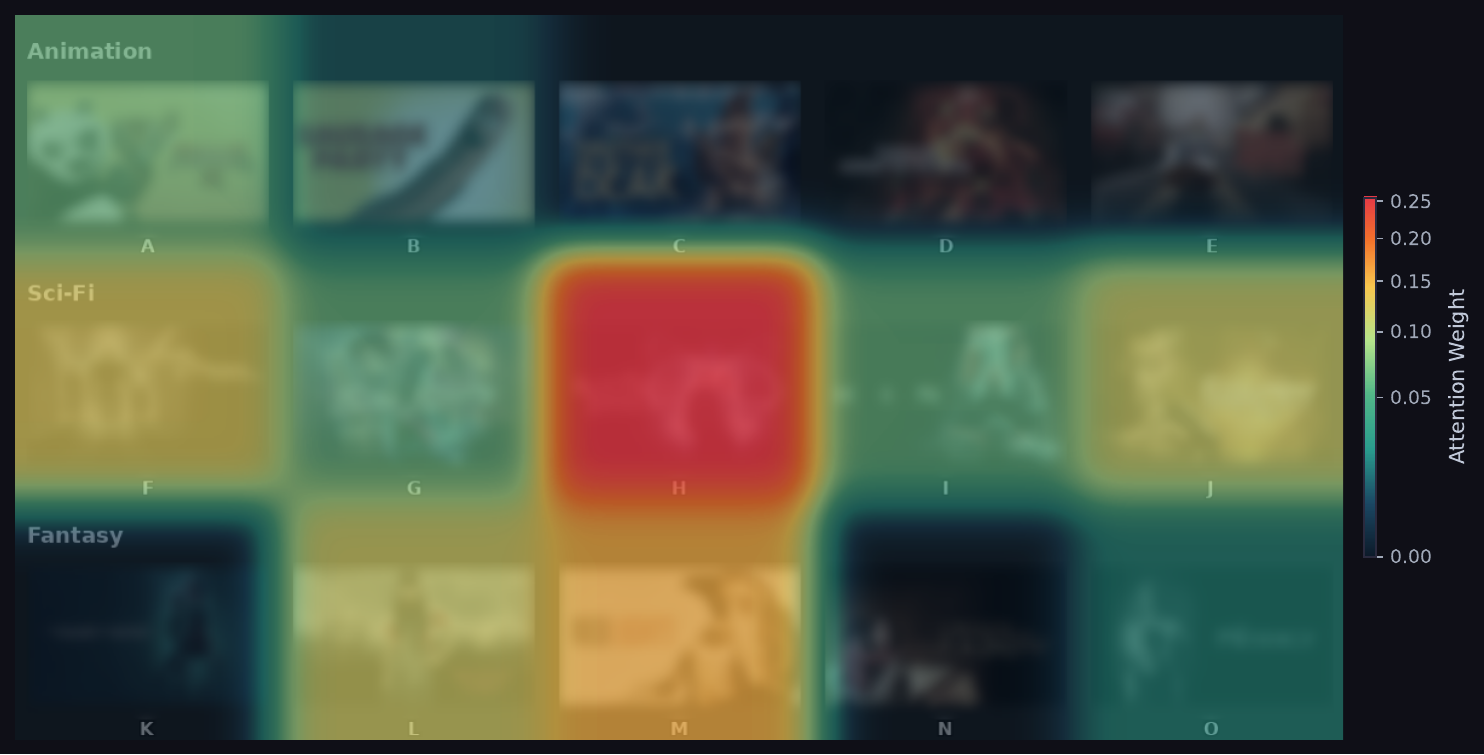}
        \caption{\backbone.}
        \label{fig:case_study_baseline}
    \end{subfigure}
    \hfill
    \begin{subfigure}[t]{0.32\linewidth}
        \centering
        \includegraphics[width=\linewidth]{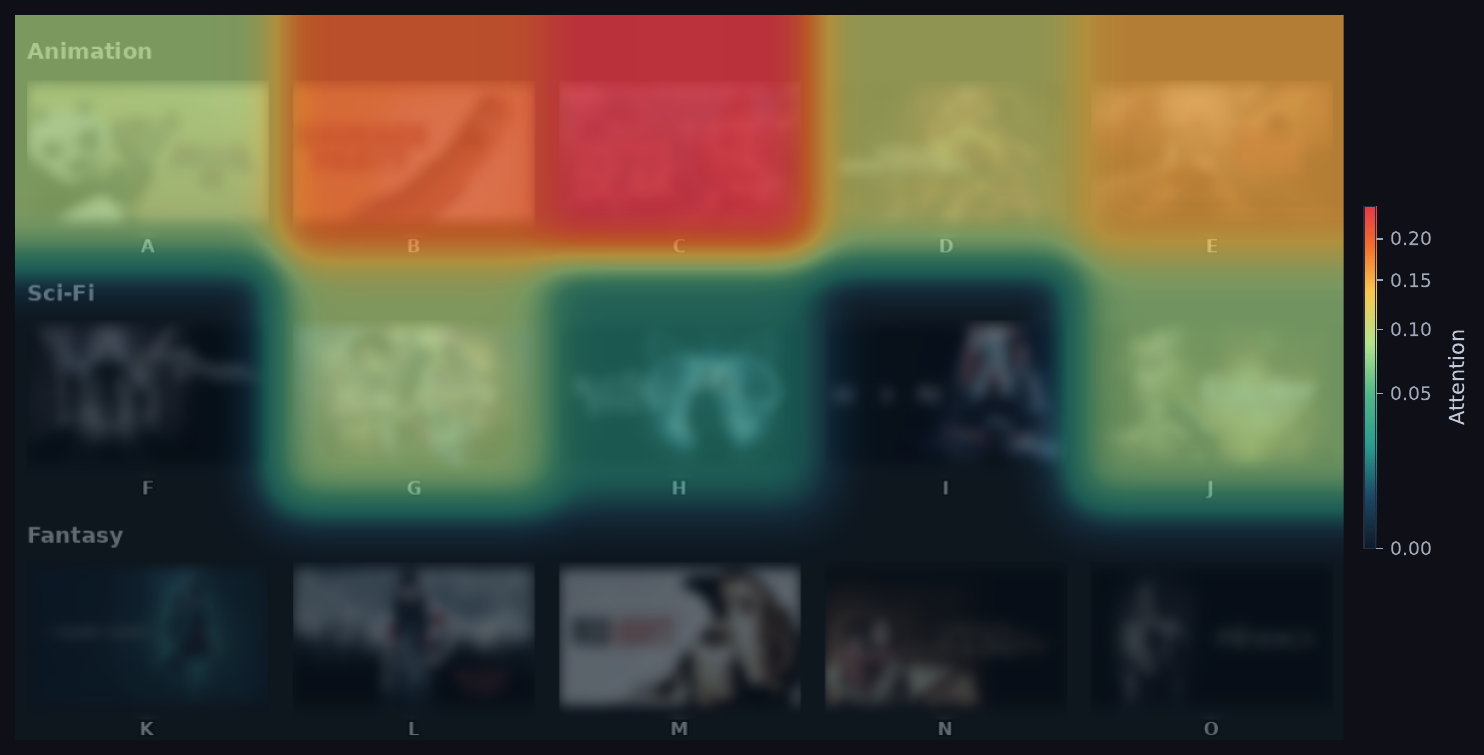}
        \caption{Human gaze.}
        \label{fig:case_study_human}
    \end{subfigure}
    \hfill
    \begin{subfigure}[t]{0.32\linewidth}
        \centering
        \includegraphics[width=\linewidth]{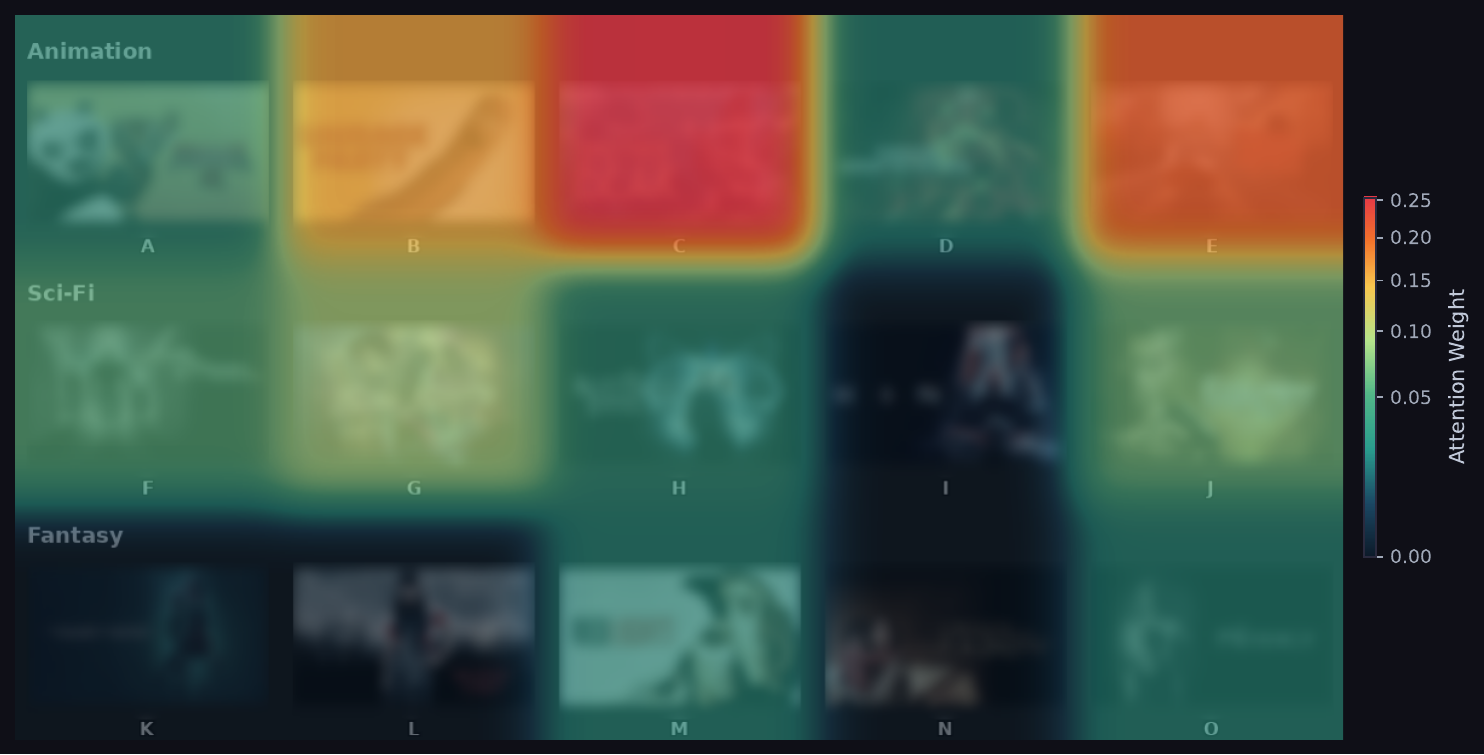}
        \caption{\sysname.}
        \label{fig:case_study_trained}
    \end{subfigure}
    \captionsetup{skip=1.3pt}
    \caption{Case study of slot-level attention on \textit{Qwen3-VL}. (a)~Original backbone attention; (b)~Human gaze; (c)~Attention from \sysname. Posters are blurred for copyright reasons.}
    \label{fig:case_study}
\end{figure}

Figure~\ref{fig:case_study} presents a qualitative comparison of slot-level attention distributions for a representative session on \textit{Qwen3-VL}, contrasting \backbone, human gaze, and \sysname. The \backbone model (Figure~\ref{fig:case_study_baseline}) exhibits a diffuse and spatially unstructured attention pattern, with relevance spread across the grid and no clear concentration on any particular slot, reflecting the absence of user-specific guidance. In contrast, the human gaze distribution (Figure~\ref{fig:case_study_human}) is sharply peaked, with fixation concentrated on a few salient slots, consistent with the gaze stability documented in Sec.~\ref{sec:motivation}. \sysname (Figure~\ref{fig:case_study_trained}) substantially narrows this gap: after personalized prompt tuning, the model's slot-level relevance shifts toward the most-attended regions of the human gaze map, while suppressing the diffuse background activation present in \backbone. Crucially, the slot receiving the highest human fixation weight is ranked first by \sysname but not by \backbone, directly corresponding to the gain in CSH@1 in Table~\ref{tab:main_results}. This alignment between model-internal relevance and human gaze is not merely cosmetic; higher top-slot alignment reliably translates into stronger click prediction accuracy, as shown in Sec.~\ref{sec:main_results}. Overall, this case study provides a concrete illustration that \textit{\textbf{personalized soft prompts redirect a frozen VLM's visual attention toward user-characteristic fixation patterns}}, validating at the instance level the perceptual alignment mechanism behind \sysname's quantitative gains.

\subsection{Generalization to Search Interfaces}
\label{sec:generalization}
To assess whether \sysname generalizes beyond the carousel-based recommendation setting, we conduct additional experiments on AdSERP~\cite{latifzadehVersatileDataset2025a}, a publicly available eye-tracking dataset collected in a search engine results page (SERP) environment. 
Unlike RecGaze, AdSERP contains vertically stacked sponsored-search pages rather than fixed-grid carousels, providing a visually distinct testbed for evaluating generalization. 
Following the same preprocessing protocol as RecGaze, we retain sessions with both fixation and click records, yielding 493 interactions from 44 users.

We use \textit{AttnLRP} on both \textit{Qwen3-VL-4B-Instruct} and \textit{InternVL3.5-4B-Instruct} for this study and report the results in Table~\ref{tab:attnlrp_results}. 
\sysname achieves consistent gains over \backbone on both VLM backbones, with particularly clear improvements in top-slot alignment and prediction quality; for example, CSH@1 and Accuracy increase by over 28\% on InternVL3.5, while Qwen3-VL shows even larger gains on top-slot overlap and click prediction.
This pattern mirrors the findings on RecGaze, showing that personalized attention alignment improves both gaze correspondence and downstream prediction quality across interface types. 
\textit{\textbf{These results provide strong evidence of the generalizability of \sysname, showing that it is not confined to carousel-based recommendation interfaces but can transfer effectively to a visually and structurally distinct sponsored-search domain.}}

\input{Tables/attnlrp_results}

%% file: Tables/prompt.tex
\begin{figure}[t]
\centering
\begin{tcolorbox}[
  colback=gray!5, colframe=gray!60,
  title=Multimodal Prompt to VLM Backbones,
  fonttitle=\bfseries\small,
  boxrule=0.5pt, arc=2pt
]

\footnotesize
\textbf{User Interface:} $\langle$\texttt{<|vision\_start|>...<|vision\_end|>}$\rangle$

\smallskip
\textbf{User Profile:} Top\_genre: Comedy, ...; Preferred\_genres: Drama, ...

\smallskip
\textbf{Instruction:} You are a sophisticated user behavior emulator. Given the user profile above,
simulate the user's final selection on this movie recommendation interface.
Respond with exactly one uppercase English letter---the letter displayed below the poster you select---and nothing else.

\end{tcolorbox}
\captionsetup{skip=-0.5pt}
\caption{Multimodal prompt template used in \textsc{FixATE}.}
\label{fig:multimodal_prompt}
\vspace{-1.8em}
\end{figure}

%% file: Tables/main_results.tex
\begin{table*}[t]
\captionsetup{skip=1.3pt}
\caption{Evaluation results across probing operators and VLM backbones. \backbone denotes the baseline without personalized soft prompts, while \sysname uses the learned user-specific prompt $\mathbf{P}_u$. Improv. denotes the relative improvement (\%).}
\label{tab:main_results}
\centering
\footnotesize
\setlength\tabcolsep{4pt}
\renewcommand{\arraystretch}{0.9}
\begin{tabular}{@{}ll c ccc ccc ccc ccc@{}}
\toprule
\multirow{2}{*}{\textbf{Model}} 
& \multirow{2}{*}{\textbf{Method}} 
& \multirow{2}{*}{\textbf{Setting}} 
& \multicolumn{9}{c}{\textbf{Attention alignment metrics}} 
& \multicolumn{3}{c}{\textbf{Prediction-level metrics}} \\
\cmidrule(lr){4-12} \cmidrule(lr){13-15}
& & 
& \textbf{CSH@1} & \textbf{CSH@3} & \textbf{CSH@5} 
& \textbf{TGO@1} & \textbf{TGO@3} & \textbf{TGO@5} 
& \textbf{Cosine} & \textbf{JS}$\downarrow$ & \textbf{KL}$\downarrow$ 
& \textbf{Accuracy} & \textbf{LogLoss}$\downarrow$ & \textbf{AUC} \\
\midrule
\multirow{9}{*}{\shortstack[l]{Qwen3-VL-\\4B-Instruct}}
 & \multirow{3}{*}{GLIMPSE}
 & \backbone & 0.1447 & 0.3947 & 0.5789 & 0.1447 & 0.3684 & 0.5211 & 0.5270 & 0.2552 & 1.1033 & 0.1053 & 2.7860 & 0.5893 \\
 & & \sysname & \textbf{0.2368} & \textbf{0.4211} & \textbf{0.6184} & \textbf{0.1974} & \textbf{0.4035} & \textbf{0.5579} & \textbf{0.5559} & \textbf{0.2354} & \textbf{0.9986} & \textbf{0.1974} & \textbf{2.6042} & \textbf{0.6607} \\
 \cdashline{3-15}
 & & \imp{Improv.} & \imp{63.64\%} & \imp{6.67\%} & \imp{6.82\%} & \imp{36.36\%} & \imp{9.52\%} & \imp{7.07\%} & \imp{5.49\%} & \imp{7.76\%} & \imp{9.49\%} & \imp{87.50\%} & \imp{6.53\%} & \implast{12.12\%} \\
\cmidrule(l){2-15}
 & \multirow{3}{*}{AttnLRP}
 & \backbone & 0.0395 & 0.3026 & 0.4474 & 0.0658 & 0.3158 & 0.4974 & 0.4828 & 0.2704 & 1.2187 & 0.1053 & 2.7860 & 0.5893 \\
 & & \sysname & \textbf{0.1316} & \textbf{0.4605} & \textbf{0.5921} & \textbf{0.1184} & \textbf{0.3860} & \textbf{0.5289} & \textbf{0.5350} & \textbf{0.2559} & \textbf{1.0954} & \textbf{0.2105} & \textbf{2.5759} & \textbf{0.6701} \\
 \cdashline{3-15}
 & & \imp{Improv.} & \imp{\textbf{\textcolor{red!70!black}{233.33\%}}} & \imp{52.17\%} & \imp{32.35\%} & \imp{80.00\%} & \imp{22.22\%} & \imp{6.35\%} & \imp{10.82\%} & \imp{5.36\%} & \imp{10.12\%} & \imp{\textbf{\textcolor{red!70!black}{100.00\%}}} & \imp{7.54\%} & \implast{13.72\%} \\
\cmidrule(l){2-15}
 & \multirow{3}{*}{\shortstack[l]{Attention\\Rollout}}
 & \backbone & 0.0921 & 0.3158 & 0.5132 & 0.0658 & 0.2939 & 0.4763 & 0.3512 & 0.3534 & 1.7114 & 0.1053 & 2.7860 & 0.5893 \\
 & & \sysname & \textbf{0.1579} & \textbf{0.3947} & \textbf{0.5263} & \textbf{0.1711} & \textbf{0.3553} & \textbf{0.5000} & \textbf{0.5400} & \textbf{0.2720} & \textbf{1.1117} & \textbf{0.1711} & \textbf{2.6208} & \textbf{0.6513} \\
 \cdashline{3-15}
 & & \imp{Improv.} & \imp{71.43\%} & \imp{25.00\%} & \imp{2.56\%} & \imp{160.00\%} & \imp{20.90\%} & \imp{4.97\%} & \imp{53.76\%} & \imp{23.03\%} & \imp{35.04\%} & \imp{62.50\%} & \imp{5.93\%} & \implast{10.52\%} \\
\midrule
\multirow{9}{*}{\shortstack[l]{InternVL3.5-\\4B-Instruct}}
 & \multirow{3}{*}{GLIMPSE}
 & \backbone & 0.0395 & 0.1447 & 0.3553 & 0.0526 & 0.1491 & 0.3474 & 0.3725 & 0.3354 & 1.5384 & 0.1184 & 2.7958 & 0.5203 \\
 & & \sysname & \textbf{0.0789} & \textbf{0.2368} & \textbf{0.3947} & \textbf{0.1053} & \textbf{0.2544} & \textbf{0.4132} & \textbf{0.4986} & \textbf{0.2882} & \textbf{1.1978} & \textbf{0.1316} & \textbf{2.6864} & \textbf{0.5724} \\
 \cdashline{3-15}
 & & \imp{Improv.} & \imp{100.00\%} & \imp{63.64\%} & \imp{11.11\%} & \imp{100.00\%} & \imp{70.59\%} & \imp{18.94\%} & \imp{33.86\%} & \imp{14.07\%} & \imp{22.14\%} & \imp{11.11\%} & \imp{3.92\%} & \implast{10.01\%} \\
\cmidrule(l){2-15}
 & \multirow{3}{*}{AttnLRP}
 & \backbone & 0.0526 & 0.1447 & 0.2763 & 0.0395 & 0.1579 & 0.3263 & 0.3384 & 0.3586 & 1.7537 & 0.1184 & 2.7958 & 0.5203 \\
 & & \sysname & \textbf{0.1053} & \textbf{0.2632} & \textbf{0.3158} & \textbf{0.1184} & \textbf{0.2325} & \textbf{0.3632} & \textbf{0.3992} & \textbf{0.3210} & \textbf{1.5118} & \textbf{0.1711} & \textbf{2.6088} & \textbf{0.6344} \\
 \cdashline{3-15}
 & & \imp{Improv.} & \imp{100.00\%} & \imp{81.82\%} & \imp{14.29\%} & \imp{200.00\%} & \imp{47.22\%} & \imp{11.29\%} & \imp{17.99\%} & \imp{10.50\%} & \imp{13.79\%} & \imp{44.44\%} & \imp{6.69\%} & \implast{21.93\%} \\
\cmidrule(l){2-15}
 & \multirow{3}{*}{\shortstack[l]{Attention\\Rollout}}
 & \backbone & 0.0395 & 0.1184 & 0.2632 & 0.0526 & 0.1447 & 0.3237 & 0.3587 & 0.3416 & 1.5791 & 0.1184 & 2.7958 & 0.5203 \\
 & & \sysname & \textbf{0.1053} & \textbf{0.1974} & \textbf{0.3684} & \textbf{0.0658} & \textbf{0.2368} & \textbf{0.3789} & \textbf{0.4881} & \textbf{0.2946} & \textbf{1.2137} & \textbf{0.1974} & \textbf{2.6759} & \textbf{0.6136} \\
 \cdashline{3-15}
 & & \imp{Improv.} & \imp{\textbf{\textcolor{red!70!black}{166.67\%}}} & \imp{66.67\%} & \imp{40.00\%} & \imp{25.00\%} & \imp{63.64\%} & \imp{17.07\%} & \imp{36.06\%} & \imp{13.76\%} & \imp{23.14\%} & \imp{\textbf{\textcolor{red!70!black}{66.67\%}}} & \imp{4.29\%} & \implast{17.93\%} \\
\bottomrule
\end{tabular}
\end{table*}

%% file: Tables/ablation.tex
\begin{table*}[t]
\captionsetup{skip=1.3pt}
\caption{Ablation study results using \textit{AttnLRP} as the probing operator. Best values per model are highlighted with a light red background and boldface; values better than \backbone are highlighted with a light blue background.}
\label{tab:ablation}
\centering
\footnotesize
\setlength\tabcolsep{4pt}
\renewcommand{\arraystretch}{1.0}
\begin{tabular}{@{}ll ccc ccc ccc ccc@{}}
\toprule
\multirow{2}{*}{\textbf{Model}} 
& \multirow{2}{*}{\textbf{Setting}} 
& \multicolumn{9}{c}{\textbf{Attention alignment metrics}} 
& \multicolumn{3}{c}{\textbf{Prediction-level metrics}} \\
\cmidrule(lr){3-11} \cmidrule(lr){12-14}
& 
& \textbf{CSH@1} & \textbf{CSH@3} & \textbf{CSH@5} 
& \textbf{TGO@1} & \textbf{TGO@3} & \textbf{TGO@5} 
& \textbf{Cosine} & \textbf{JS}$\downarrow$ & \textbf{KL}$\downarrow$ 
& \textbf{Accuracy} & \textbf{LogLoss}$\downarrow$ & \textbf{AUC} \\
\midrule
\multirow{7}{*}{\shortstack[l]{Qwen3-VL-\\4B-Instruct}}
 & \backbone 
 & 0.0395 & 0.3026 & 0.4474 & 0.0658 & 0.3158 & 0.4974 & 0.4828 & 0.2704 & 1.2187 & 0.1053 & 2.7860 & 0.5893 \\
 & Rand.~SP 
 & \bettercell{0.0658} & {0.1711} & {0.3684} & {0.0526} & {0.1263} & {0.4263} & {0.4250} & {0.3055} & {1.3509} & {0.0789} & {2.8992} & {0.5620} \\
 & $\boldsymbol{z}_u = \mathbf{1}$
 & {0.0263} & \bettercell{0.3289} & \bettercell{0.5263} & \bettercell{0.0789} & {0.3070} & {0.4816} & \bettercell{0.4964} & {0.2727} & \bettercell{1.1892} & {0.0921} & {2.8945} & {0.5883} \\
 & w/o $\omega_n$
 & \bettercell{0.0789} & \bettercell{0.3158} & \bettercell{0.4737} & \bettercell{0.0921} & \bettercell{0.3246} & \bettercell{0.5026} & \bettercell{0.4890} & \bettercell{0.2690} & \bettercell{1.1926} & \bettercell{0.1184} & \bettercell{2.7594} & \bettercell{0.6118} \\
 & w/o $\mathcal{L}_{\text{Attn}}$ 
 & 0.0395 & {0.2500} & \bettercell{0.4605} & \bettercell{0.0789} & {0.2675} & {0.4211} & {0.4336} & {0.2983} & {1.3272} & {0.0921} & {2.8066} & {0.5780} \\
 & w/o $\mathcal{L}_{\text{NTP}}$ 
 & \bettercell{0.1184} & \bettercell{0.3553} & \bettercell{0.5526} & \bestcell{0.1447} & \bettercell{0.3728} & \bestcell{0.5711} & \bettercell{0.5092} & \bettercell{0.2658} & \bettercell{1.1765} & \bettercell{0.1316} & \bettercell{2.7044} & \bettercell{0.6335} \\
 & \sysname 
 & \bestcell{0.1316} & \bestcell{0.4605} & \bestcell{0.5921} & \bettercell{0.1184} & \bestcell{0.3860} & \bettercell{0.5289} & \bestcell{0.5350} & \bestcell{0.2559} & \bestcell{1.0954} & \bestcell{0.2105} & \bestcell{2.5759} & \bestcell{0.6701} \\
\midrule
\multirow{7}{*}{\shortstack[l]{InternVL3.5-\\4B-Instruct}}
 & \backbone 
 & 0.0526 & 0.1447 & 0.2763 & 0.0395 & 0.1579 & 0.3263 & 0.3384 & 0.3586 & 1.7537 & 0.1184 & 2.7958 & 0.5203 \\
 & Rand.~SP 
 & {0.0263} & {0.1184} & {0.2237} & {0.0132} & \bettercell{0.1754} & {0.3053} & {0.3254} & {0.3617} & {1.7702} & {0.0658} & {3.2524} & {0.4596} \\
 & $\boldsymbol{z}_u = \mathbf{1}$
 & {0.0395} & \bettercell{0.2368} & \bettercell{0.2895} & {0.0263} & \bettercell{0.2061} & {0.3211} & \bettercell{0.3463} & \bettercell{0.3551} & \bettercell{1.7096} & {0.0789} & {3.1114} & {0.5078} \\
 & w/o $\omega_n$
 & \bettercell{0.0658} & \bettercell{0.1579} & 0.2763 & \bettercell{0.0526} & \bettercell{0.1798} & \bettercell{0.3316} & {0.3361} & \bettercell{0.3555} & \bettercell{1.6988} & \bettercell{0.1184} & {2.8311} & \bettercell{0.5290} \\
 & w/o $\mathcal{L}_{\text{Attn}}$ 
 & {0.0263} & {0.1053} & {0.2500} & 0.0395 & \bettercell{0.1667} & {0.3158} & {0.3365} & \bettercell{0.3534} & \bettercell{1.6567} & {0.1053} & {3.1566} & {0.4727} \\
 & w/o $\mathcal{L}_{\text{NTP}}$ 
 & \bettercell{0.0658} & \bettercell{0.2105} & \bestcell{0.4079} & \bettercell{0.0789} & \bestcell{0.2368} & \bestcell{0.3763} & \bettercell{0.3746} & \bettercell{0.3283} & \bestcell{1.4573} & \bettercell{0.1316} & \bettercell{2.7927} & \bettercell{0.5639} \\
 & \sysname 
 & \bestcell{0.1053} & \bestcell{0.2632} & \bettercell{0.3158} & \bestcell{0.1184} & \bettercell{0.2325} & \bettercell{0.3632} & \bestcell{0.3992} & \bestcell{0.3210} & \bettercell{1.5118} & \bestcell{0.1711} & \bestcell{2.6088} & \bestcell{0.6344} \\
\bottomrule
\end{tabular}
\end{table*}

%% file: Tables/attnlrp_results.tex
\begin{table}[t]
\captionsetup{skip=1.3pt}
\caption{Evaluation of \textit{AttnLRP} on the \textit{AdSERP} dataset.}
\label{tab:attnlrp_results}
\centering
\scriptsize
\setlength\tabcolsep{2pt}
\renewcommand{\arraystretch}{0.9}
\resizebox{\columnwidth}{!}{%
\begin{tabular}{@{}l c cccccc@{}}
\toprule
\textbf{Model} & \textbf{Setting}
& \textbf{CSH@1} & \textbf{TGO@1} & \textbf{KL}$\downarrow$ 
& \textbf{Accuracy} & \textbf{LogLoss}$\downarrow$ & \textbf{AUC} \\
\midrule
\multirow{3}{*}{\shortstack[l]{Qwen3-VL-\\4B-Instruct}}
 & \backbone & 0.2727 & 0.3939 & 0.5797 & 0.2727 & 1.4450 & 0.6760 \\
 & \sysname & \textbf{0.3636} & \textbf{0.5758} & \textbf{0.5547} & \textbf{0.3636} & \textbf{1.3516} & \textbf{0.6851} \\
 \cdashline{2-8}
 & Improv. & \imp{33.33\%} & \imp{46.15\%} & \imp{4.31\%} & \imp{33.33\%} & \imp{6.46\%} & \implast{1.35\%} \\
\midrule
\multirow{3}{*}{\shortstack[l]{InternVL3.5-\\4B-Instruct}}
 & \backbone & 0.2121 & 0.4242 & 0.8482 & 0.2121 & 1.8929 & 0.5048 \\
 & \sysname & \textbf{0.2727} & \textbf{0.4848} & \textbf{0.7355} & \textbf{0.2727} & \textbf{1.7947} & \textbf{0.5547} \\
 \cdashline{2-8}
 & Improv. & \imp{28.57\%} & \imp{14.29\%} & \imp{13.29\%} & \imp{28.57\%} & \imp{5.19\%} & \implast{9.89\%} \\
\bottomrule
\end{tabular}%
}
\vspace{-1em}
\end{table}

%% file: Main/Related_works.tex
\section{Related Work}


\textbf{LLM-based User Simulators for Recommender Systems.}
Recent work has explored LLM agents as user simulators for recommender systems, typically following a perception--memory--action architecture~\cite{pengSurveyLLMpowered2025}, which offers a natural paradigm for modeling sequential user decision-making. Representative systems such as Agent4Rec~\cite{zhangGenerativeAgents2024} and RecAgent~\cite{wangUserBehavior2025} construct generative agents with profile, memory, and action modules, while subsequent work has further improved simulation quality through preference modeling~\cite{zhangLLMpoweredUser2025}, evaluation confound analysis~\cite{zhuHowReliable2024}, and cross-domain preference transfer~\cite{chenLassoLarge2025}, thereby improving both simulation fidelity and applicability. Despite this progress, all existing simulators operate in a text-only modality~\cite{bougieSimUSERSimulating2025, chenLassoLarge2025, zhangLLMpoweredUser2025}, reasoning over item metadata rather than perceiving the visual interface and layout structure through which real users browse recommendations. Our work addresses this gap by introducing a VLM-based simulator that perceives rendered recommendation interfaces as images, bringing user simulation closer to the real browsing experience.

\textbf{Gaze-Guided Visual Attention Alignment for VLMs.}
A growing body of work seeks to align VLM attention with human gaze through
three main approaches: input and prompt editing~\cite{wanContrastiveRegion2024,
yuAttentionPrompting2024, zhuGuidingMedical2025}, which modifies the image or
prompt before it reaches the VLM; internal attention reshaping~\cite{paniGazeVLMBridging2025},
which regularizes transformer attention maps against gaze heatmaps during training;
and gaze-based training objectives~\cite{yanVoilaAAligning2024}, which integrate
gaze signals directly into architectural components.

A critical limitation unifies all these methods: they operate at the
\textbf{task level or population-averaged level}, treating human attention as a
universal signal. Yet individual gaze behavior varies meaningfully across
users~\cite{strohmLearningUser2024, gonzalezGenderawareSaliency2025},
suggesting that simple group- or task-level alignment is insufficient.
Our work is the first to introduce \textbf{personalized, user-level} gaze
alignment for VLMs.

%% file: Main/Conclusion.tex
\section{Conclusion, Limitations, and Future Work}

We proposed \sysname, a framework that steers a frozen VLM's visual attention toward each user's characteristic gaze pattern through gaze-guided attention probing and a factorized soft prompt design. Empirical analysis confirmed that personalized gaze patterns are strongly predictive of click behavior and are lost under text-based interface processing. Experiments across three probing operators and two VLM backbones showed consistent improvements in both attention alignment and click prediction, with ablation studies identifying personalized gaze supervision as the primary driver. 

Several limitations point to future directions. \textbf{First,} real-world eye-tracking data in recommendation settings remains scarce, constraining the scale and diversity of our evaluation; broader validation will become feasible as more gaze datasets are available. \textbf{Second,} collecting per-user eye-tracking data is costly and impractical at scale; in such cases, low-cost and readily accessible proxies such as cursor traces warrant further investigation as substitutes for real human fixation. \textbf{Third,} the dataset used lacks temporal links across sessions, preventing our framework from modeling how user attention evolves over time; incorporating longitudinal histories could further enrich the learned attentional priors.

%% file: Main/Supplementary.tex
\section{Statistical Significance of Experimental Results}
\label{sec:supp_significance}

\input{Tables/supp_recgaze}
\input{Tables/supp_adserp}

All results reported in the main paper (Tables \ref{tab:main_results} and~\ref{tab:attnlrp_results}) are averaged over 5 independent runs with different random seeds.
To rigorously assess the reliability of \sysname's improvements, we conduct  $t$-tests between \sysname and \backbone for every (Model $\times$ Probing Operator $\times$ Metric) combination.
Tables~\ref{tab:supp_recgaze} and~\ref{tab:supp_adserp} present the extended results with standard deviations and significance markers.

\subsection{Extended Results with Standard Deviations}
\label{sec:supp_extended}

Tables \ref{tab:supp_recgaze} and \ref{tab:supp_adserp} extend the main Tables \ref{tab:main_results} and \ref{tab:attnlrp_results} with standard deviations and statistical significance markers for a representative subset of metrics.
We select six metrics that span both evaluation dimensions: CSH@1 and TGO@1 (top-slot attention alignment), KL$\downarrow$ (global distributional alignment), and Accuracy, LogLoss$\downarrow$, AUC (prediction quality).

\subsection{Discussion}
\label{sec:supp_discussion}

Several observations emerge from the extended statistical analysis.
\paragraph{The improvements are statistically significant across the board.}
Across all comparisons in Table~\ref{tab:supp_recgaze}, every single one reaches at least $p<0.05$, and the large majority achieve $p<0.01$.
The same pattern holds in Table~\ref{tab:supp_adserp}, where 11 out of 12 comparisons are significant.
This confirms that the gains reported in the main paper are robust and unlikely to arise from random seed variation.
\paragraph{Prediction-level metrics exhibit particularly strong significance.}
AUC, LogLoss, and Accuracy achieve $p<0.01$ in nearly all settings on both datasets.
These metrics aggregate over the full test set, providing sufficient statistical power despite the modest dataset scale.
The consistently low standard deviations on AUC further indicate that \sysname's behavioral improvements are stable across runs.
\paragraph{Top-slot metrics show higher variance but remain significant.}
Discrete metrics such as CSH@1 and TGO@1 naturally exhibit larger relative standard deviations, because a single test session flipping between correct and incorrect identification can shift the metric noticeably.
Despite this, all CSH@1 and TGO@1 comparisons are significant, with several reaching $p<0.01$.
This suggests that personalized fixation alignment provides a reliable and consistent benefit in identifying the user's most salient slot.
\paragraph{One exception: KL on \textit{AdSERP} (\textit{Qwen3-VL})}
The KL divergence improvement for \textit{Qwen3-VL} on \textit{AdSERP} (4.31\% relative improvement) does not reach $p<0.05$, though the direction of improvement is consistent across all 5 seeds.
We attribute this to the already low baseline KL value (0.5797), which leaves limited room for further distributional realignment.
Notably, top-slot and prediction-level metrics in the same setting are all significant, suggesting that the model still achieves meaningful behavioral alignment even when the global distributional shift is modest.

\section{Detailed Formulation of Probing Operators}
\label{sec:supp_probing}

Section~\ref{sec:probing} of the main paper defines a general probing function $\mathbf{R}=f(\mathcal{M},\mathbf{C},\mathbf{y})$ and states that it can be instantiated with three representative interpretability methods. This section provides unified mathematical formulations for all three operators, clarifying how they differ in per-layer relevance extraction and cross-layer propagation. Our implementations follow the original formulations proposed in \textit{Attention Rollout}~\cite{abnarQuantifyingAttention2020}, \textit{AttnLRP}~\cite{achtibatAttnLRPAttentionaware2024}, and \textit{GLIMPSE}~\cite{shenGLIMPSEHolistic2025}, with engineering adaptations to accommodate the multimodal input format (soft prompt + vision tokens + text tokens) and the slot-level aggregation required by our framework.

\subsection{Notation}
\label{sec:supp_probing_notation}

Consider a transformer with $L$ layers, each containing $H$ attention heads.
Let $\mathbf{A}^{(l,h)}\!\in\!\mathbb{R}^{N\times N}$ denote the attention weight matrix at layer~$l$, head~$h$, where $N$ is the total sequence length (including soft prompt tokens).
For gradient-based operators, let $s$ denote a target scalar derived from the model's output logits at the answer position (Sec.~\ref{sec:probing}), and let $\mathbf{G}^{(l,h)}=\partial s/\partial\mathbf{A}^{(l,h)}$ denote the corresponding gradient.
All three operators share a two-stage structure:
\textbf{(1)}~computing a per-layer relevance matrix $\mathbf{E}_l\!\in\!\mathbb{R}^{N\times N}$, and
\textbf{(2)}~propagating relevance across layers to obtain a global relevance matrix $\mathbf{R}\!\in\!\mathbb{R}^{N\times N}$.
The final slot-level attention is then extracted from the row of $\mathbf{R}$ corresponding to the answer token position, restricted to visual token columns, and aggregated to slots as described in Sec.~\ref{sec:probing}.

\subsection{Stage 1: Per-Layer Relevance}
\label{sec:supp_stage1}

\paragraph{Attention Rollout}
This operator uses attention weights alone without any gradient information.
The per-layer relevance is the head-averaged attention matrix with a residual identity connection and row normalization:
\begin{equation}
    \mathbf{E}_l = \mathrm{RowNorm}\!\left(\,\frac{1}{2}\cdot\frac{1}{H}\sum_{h=1}^{H}\mathbf{A}^{(l,h)} + \frac{1}{2}\cdot\mathbf{I}\,\right),
    \label{eq:rollout_layer}
\end{equation}
where $\mathrm{RowNorm}(\cdot)$ normalizes each row to sum to~1. The residual term $\tfrac{1}{2}\mathbf{I}$ accounts for the skip connections in the transformer architecture.

\paragraph{AttnLRP}
This operator modulates attention weights by their gradients and applies a sign-preserving normalization.
For each head, an element-wise product captures the gradient-weighted attention:
\begin{equation}
    \mathbf{W}^{(l,h)} = \mathbf{A}^{(l,h)}\odot\mathbf{G}^{(l,h)}.
    \label{eq:attnlrp_W}
\end{equation}
Each row is then normalized by its signed sum:
\begin{equation}
    \hat{\mathbf{E}}^{(l,h)}_{i,:} = \frac{\mathbf{W}^{(l,h)}_{i,:}}{\sum_{j}\mathbf{W}^{(l,h)}_{i,j} \;+\; \epsilon\cdot\mathrm{sign}\!\left(\sum_{j}\mathbf{W}^{(l,h)}_{i,j}\right)},
    \label{eq:attnlrp_norm}
\end{equation}
where $\epsilon>0$ ensures numerical stability.
The per-layer relevance is the mean over heads:
\begin{equation}
    \mathbf{E}_l = \frac{1}{H}\sum_{h=1}^{H}\hat{\mathbf{E}}^{(l,h)}.
    \label{eq:attnlrp_layer}
\end{equation}

\paragraph{GLIMPSE}
This operator introduces adaptive head weighting based on gradient-attention alignment.
First, a non-negative relevance tensor is computed:
\begin{equation}
    \hat{\mathbf{G}}^{(l,h)} = \mathrm{ReLU}\!\left(\mathbf{G}^{(l,h)}\odot\mathbf{A}^{(l,h)}\right).
    \label{eq:glimpse_G}
\end{equation}
Each head receives an importance score measuring the fraction of gradient-aligned attention:
\begin{equation}
    \sigma_h = \frac{\sum_{i,j}\hat{\mathbf{G}}^{(l,h)}_{i,j}}{\sum_{i,j}\mathrm{ReLU}\!\left(\mathbf{G}^{(l,h)}\right)_{i,j}+\epsilon}.
    \label{eq:glimpse_score}
\end{equation}
Head weights are obtained via temperature-scaled softmax:
\begin{equation}
    w_h = \mathrm{softmax}\!\left(\boldsymbol{\sigma}/\tau_{\mathrm{head}}\right)_h.
    \label{eq:glimpse_head_weight}
\end{equation}
The per-layer relevance is the weighted combination followed by row normalization:
\begin{equation}
    \mathbf{E}_l = \mathrm{RowNorm}\!\left(\sum_{h=1}^{H}w_h\cdot\hat{\mathbf{G}}^{(l,h)}\right).
    \label{eq:glimpse_layer}
\end{equation}

\subsection{Stage 2: Cross-Layer Propagation}
\label{sec:supp_stage2}

The three operators differ substantially in how per-layer relevance matrices are combined into a global relevance matrix~$\mathbf{R}$.

\paragraph{Attention Rollout}
uses a sequential matrix product across all layers:
\begin{equation}
    \mathbf{R} = \mathbf{E}_L\cdot\mathbf{E}_{L-1}\cdots\mathbf{E}_1.
    \label{eq:rollout_prop}
\end{equation}
This formulation traces information flow through the full transformer depth by chaining the residual-augmented attention matrices.

\paragraph{AttnLRP}
uses a depth-weighted additive aggregation:
\begin{equation}
    \mathbf{R} = \mathbf{I} + \sum_{l=1}^{L}\beta_l\cdot\mathbf{E}_l,\qquad\beta_l = \frac{l}{\sum_{l'=1}^{L}l'}.
    \label{eq:attnlrp_prop}
\end{equation}
The linear depth weights $\beta_l$ assign increasing importance to deeper layers, reflecting the intuition that later layers encode more task-relevant representations.
Crucially, this formulation is purely additive (no inter-layer matrix products), which yields greater numerical stability.

\paragraph{GLIMPSE}
uses an iterative gradient-adaptive propagation.
Layer-level importance is determined by combining gradient magnitude with a depth-scaled prior:
\begin{equation}
    \alpha_l \propto \left(\textstyle\sum\left|\mathbf{G}^{(l)}\right|\right)\cdot\mathrm{softmax}\!\left(\tau_{\mathrm{depth}}\cdot l\right)_l,
    \label{eq:glimpse_alpha}
\end{equation}
where $\mathbf{G}^{(l)}$ denotes the concatenated gradients across all heads at layer~$l$, and $\tau_{\mathrm{depth}}$ is a depth temperature.
The global relevance is computed iteratively:
\begin{equation}
    \mathbf{R}^{(0)}=\mathbf{I},\qquad\mathbf{R}^{(l)}=\mathbf{R}^{(l-1)}+\alpha_l\cdot\mathbf{E}_l\cdot\mathbf{R}^{(l-1)},\qquad\mathbf{R}=\mathbf{R}^{(L)}.
    \label{eq:glimpse_prop}
\end{equation}
Unlike \textit{AttnLRP}'s additive scheme, \textit{GLIMPSE}'s update involves a matrix product $\mathbf{E}_l\cdot\mathbf{R}^{(l-1)}$ at each step, allowing cross-layer interactions to compound. The gradient-adaptive weights~$\alpha_l$ enable the propagation to emphasize layers that are most informative for the specific output.

\subsection{Summary of Differences}
\label{sec:supp_probing_summary}

Table~\ref{tab:probing_comparison} summarizes the key design differences among the three probing operators. All three produce a global relevance matrix $\mathbf{R}$ of the same shape, from which slot-level attention is extracted identically via the procedure described in Sec.~3.2 of the main paper.

\begin{table}[t]
\centering
\caption{Comparison of the three probing operators.}
\label{tab:probing_comparison}
\resizebox{\columnwidth}{!}{%
\begin{tabular}{lccc}
\toprule
\textbf{Aspect} & \textbf{Attn Rollout} & \textbf{AttnLRP} & \textbf{GLIMPSE} \\
\midrule
Gradient required        & No  & Yes & Yes \\
Head aggregation         & Uniform mean & Uniform mean & Adaptive softmax \\
Per-layer normalization  & Row-norm with $\tfrac{1}{2}\mathbf{I}$ & Signed-sum & Row-norm after ReLU \\
Cross-layer propagation  & Sequential product & Depth-weighted additive & Iterative multiplicative \\
Differentiable           & No (post-hoc) & Yes & Yes \\
Relative cost            & Lowest & Moderate & Highest \\
\bottomrule
\end{tabular}%
}
\end{table}

\section{Extended Qualitative Analysis}
\label{sec:supp_qualitative}

\begin{figure*}[t]
    \centering
    \small

    \noindent\textbf{KINIT\_49 / task~17} (representative success)\par
    \vspace{0.35em}
    \begin{subfigure}[t]{0.31\textwidth}
        \centering
        \includegraphics[width=\linewidth]{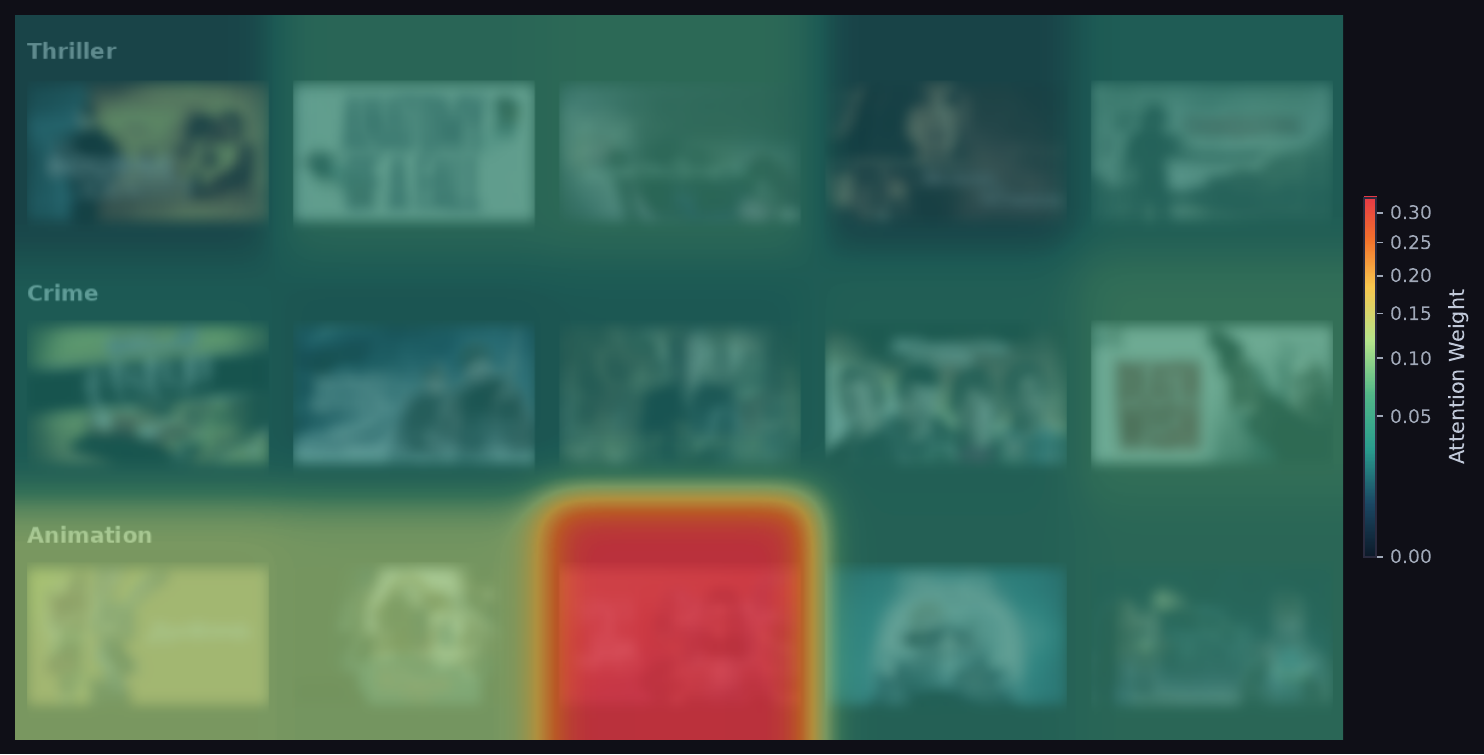}
        \caption{\backbone.}
        \label{fig:supp_sub_a}
    \end{subfigure}
    \hfill
    \begin{subfigure}[t]{0.31\textwidth}
        \centering
        \includegraphics[width=\linewidth]{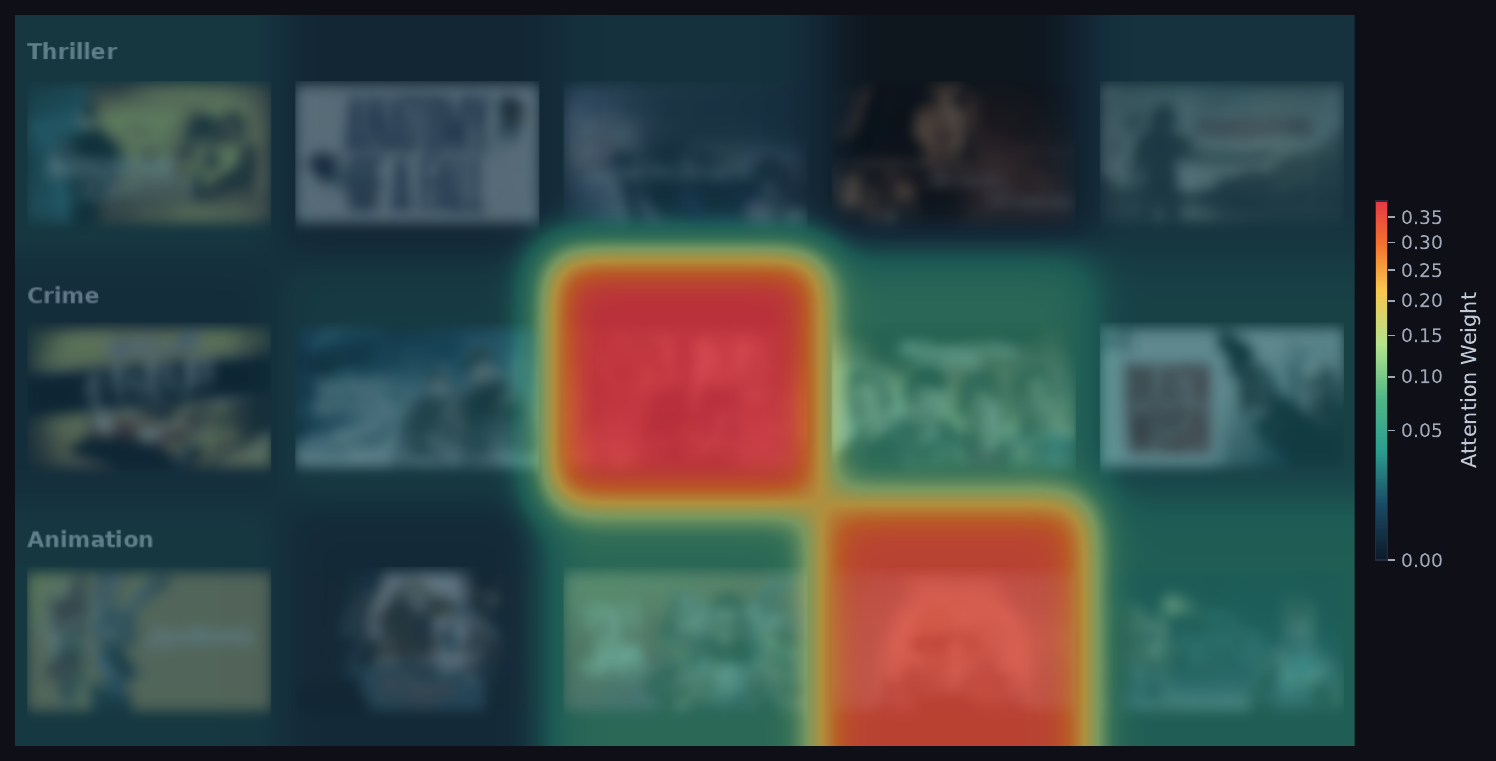}
        \caption{Human gaze.}
        \label{fig:supp_sub_b}
    \end{subfigure}
    \hfill
    \begin{subfigure}[t]{0.31\textwidth}
        \centering
        \includegraphics[width=\linewidth]{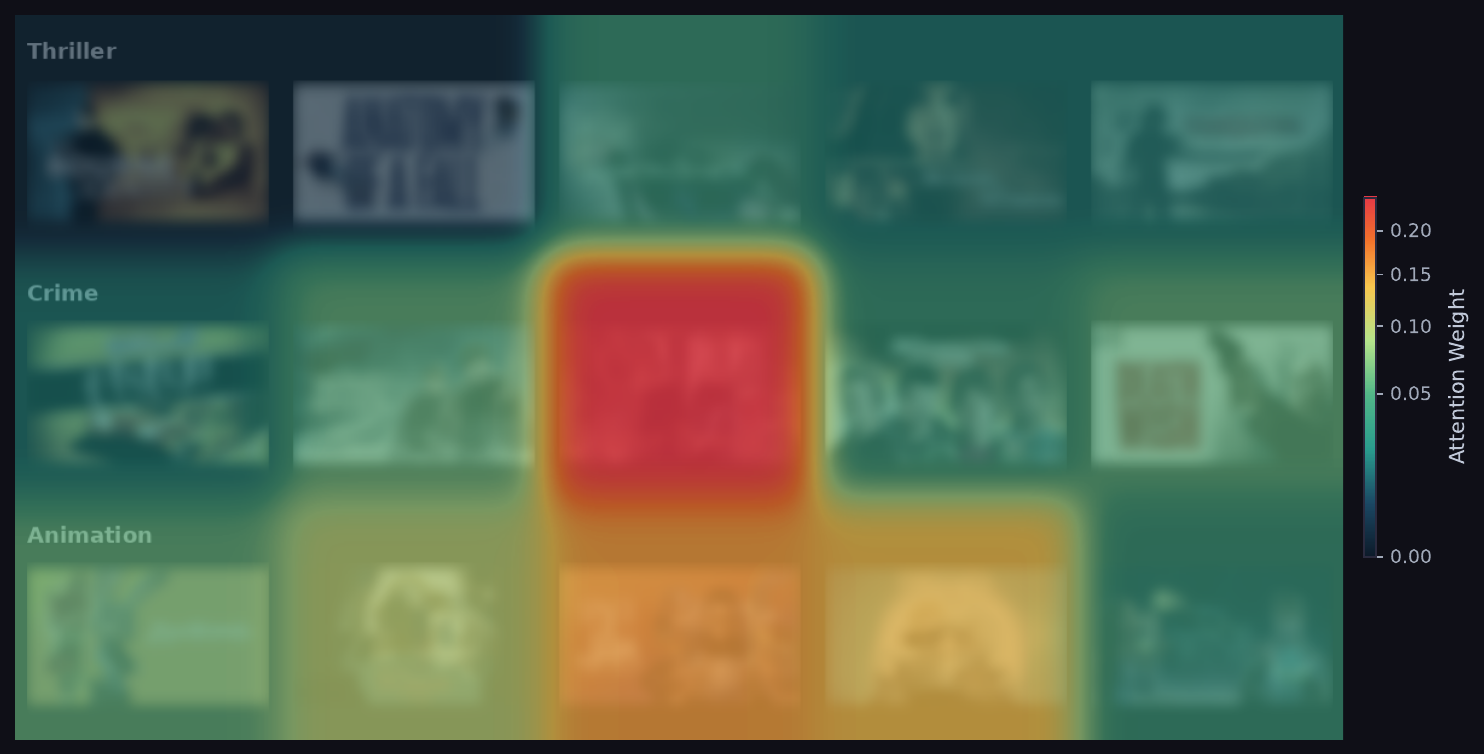}
        \caption{\sysname.}
        \label{fig:supp_sub_c}
    \end{subfigure}

    \medskip
    \noindent\textbf{UvA\_25 / task~06} (representative success)\par
    \vspace{0.35em}
    \begin{subfigure}[t]{0.31\textwidth}
        \centering
        \includegraphics[width=\linewidth]{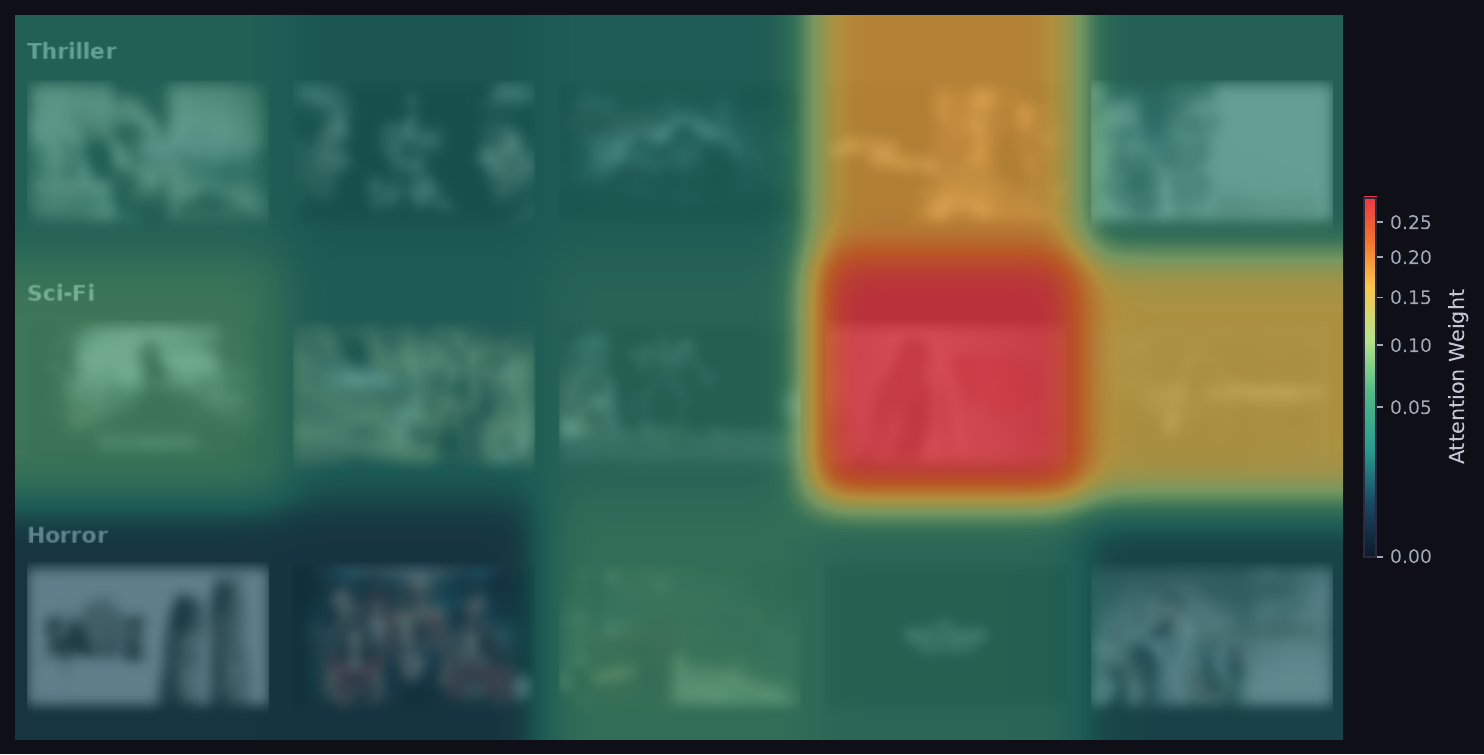}
        \caption{\backbone.}
        \label{fig:supp_sub_d}
    \end{subfigure}
    \hfill
    \begin{subfigure}[t]{0.31\textwidth}
        \centering
        \includegraphics[width=\linewidth]{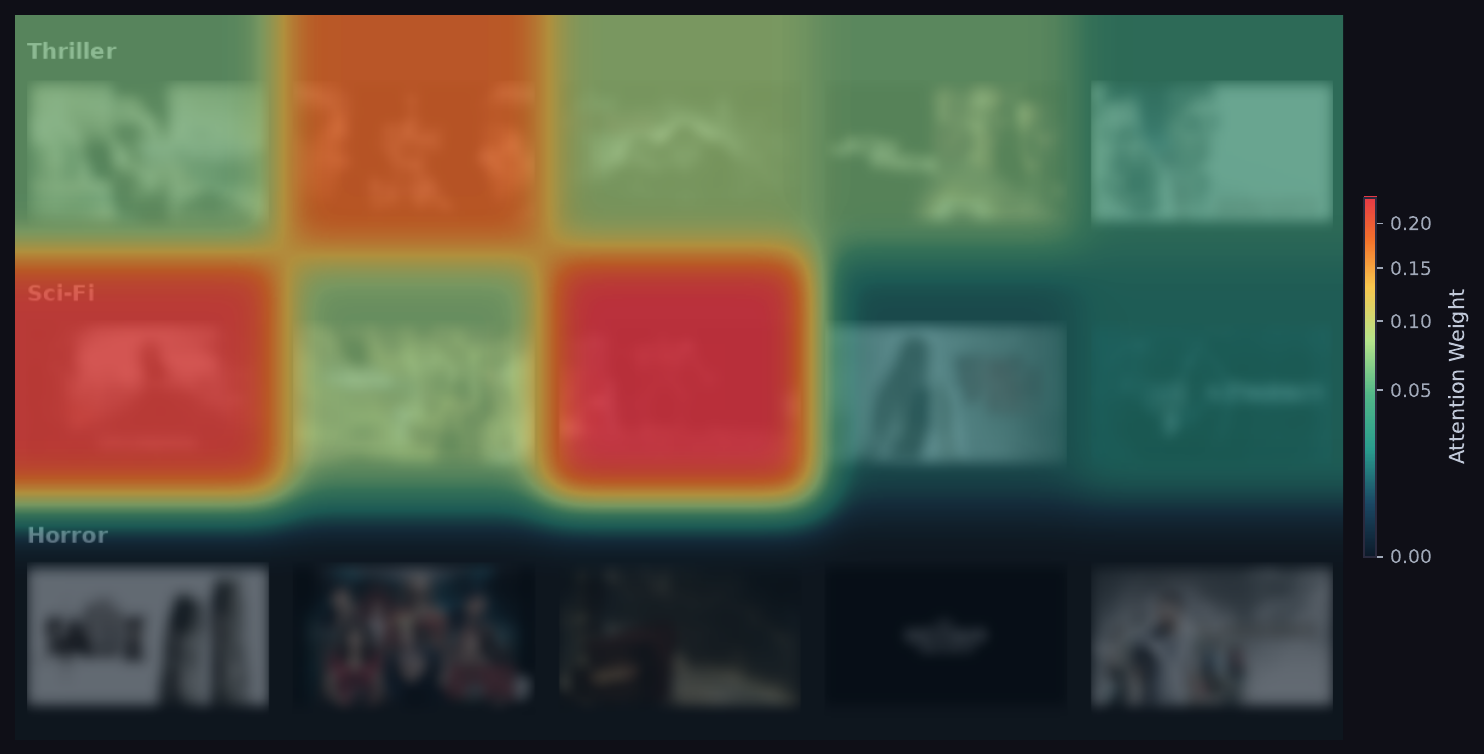}
        \caption{Human gaze.}
        \label{fig:supp_sub_e}
    \end{subfigure}
    \hfill
    \begin{subfigure}[t]{0.31\textwidth}
        \centering
        \includegraphics[width=\linewidth]{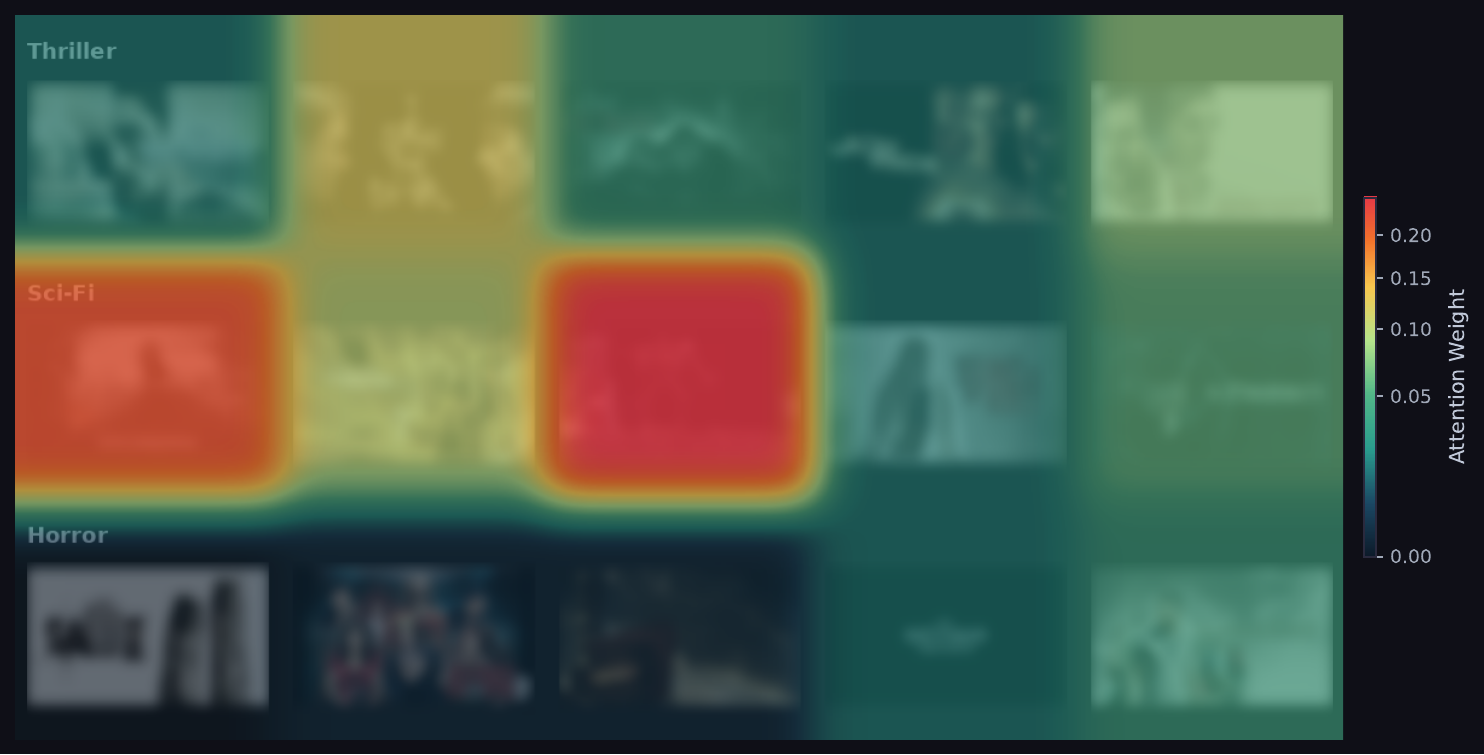}
        \caption{\sysname.}
        \label{fig:supp_sub_f}
    \end{subfigure}

    \medskip
    \noindent\textbf{KINIT\_21 / task~15} (counter-example)\par
    \vspace{0.35em}
    \begin{subfigure}[t]{0.31\textwidth}
        \centering
        \includegraphics[width=\linewidth]{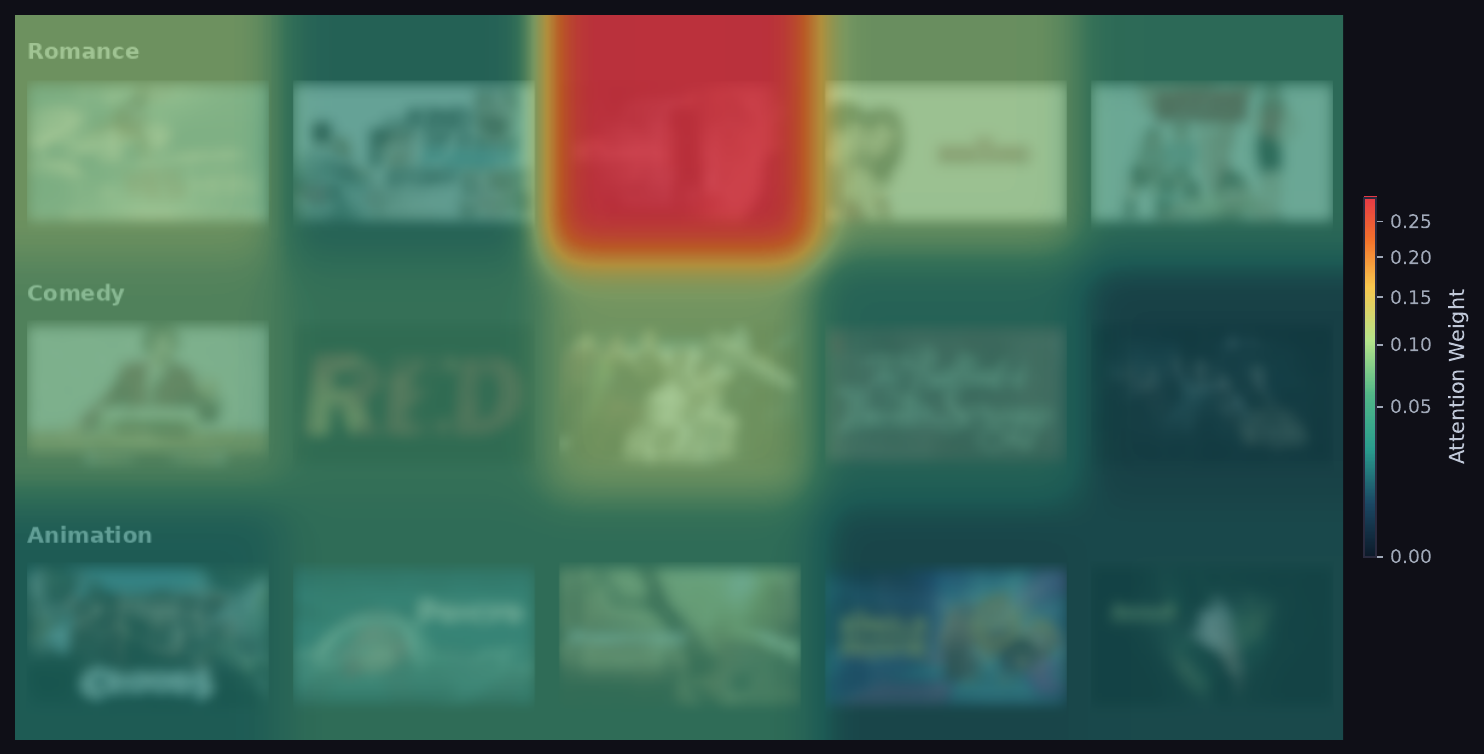}
        \caption{\backbone.}
        \label{fig:supp_sub_g}
    \end{subfigure}
    \hfill
    \begin{subfigure}[t]{0.31\textwidth}
        \centering
        \includegraphics[width=\linewidth]{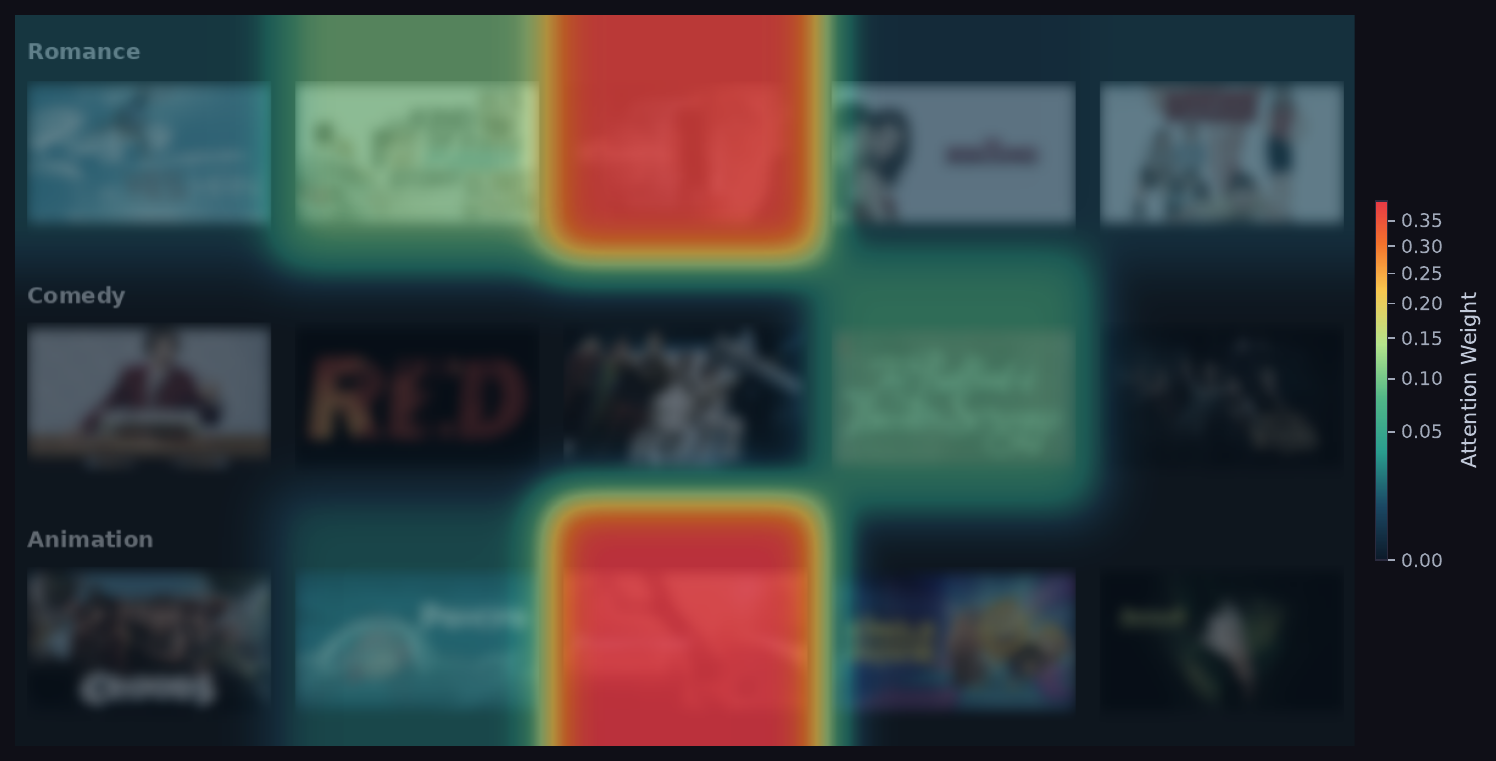}
        \caption{Human gaze.}
        \label{fig:supp_sub_h}
    \end{subfigure}
    \hfill
    \begin{subfigure}[t]{0.31\textwidth}
        \centering
        \includegraphics[width=\linewidth]{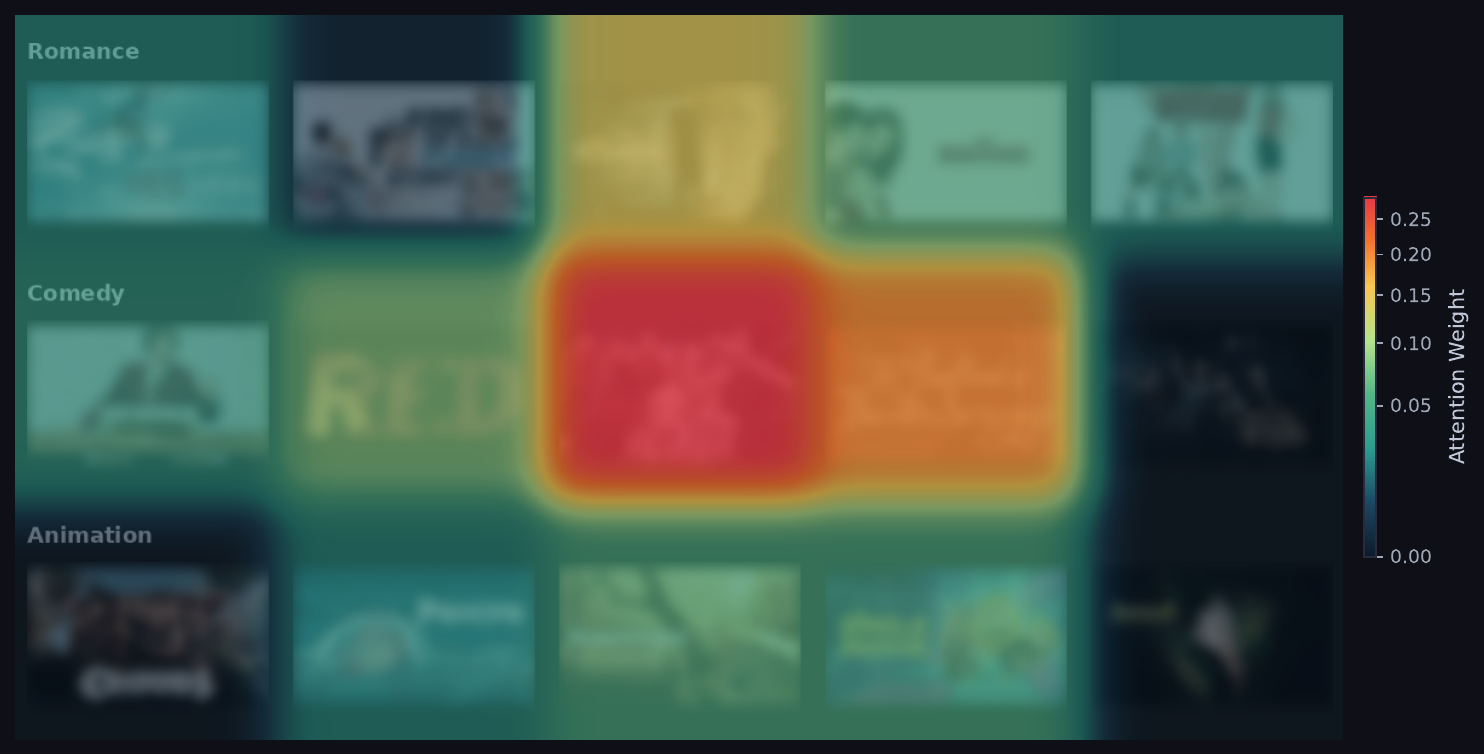}
        \caption{\sysname.}
        \label{fig:supp_sub_i}
    \end{subfigure}

    \captionsetup{skip=1.3pt}
    \caption{Supplementary qualitative comparison of slot-level attention. Each row is one session (user $\times$ task); columns are frozen \textit{Qwen3-VL} (\backbone), human fixation, and \sysname after personalized tuning. Posters are blurred where required for copyright.}
    \label{fig:supp_attention_examples}
\end{figure*}

The main paper presents a single case study (Figure~\ref{fig:case_study}) to illustrate how \sysname reshapes a frozen VLM's slot-level attention. To provide a more comprehensive picture---including both representative successes and informative failure modes---we present three additional sessions in Figure~\ref{fig:supp_attention_examples}. All visualizations use \textit{Qwen3-VL} with \textit{AttnLRP}. Each row corresponds to one test session (user $\times$ task); columns show \backbone attention, human fixation, and \sysname attention, respectively. User profile metadata from \textit{RecGaze} is listed below for reference.

\begin{itemize}
    \item \textbf{KINIT\_49 (task 17):} Top genre: Sci-Fi; Preferred genres: Action, Animation, Fantasy, Sci-Fi.
    \item \textbf{UvA\_25 (task 06):} Top genre: Crime; Preferred genres: Action, Animation, Comedy, Crime, Fantasy, Sci-Fi.
    \item \textbf{KINIT\_21 (task 15):} Top genre: Drama; Preferred genres: Action, Comedy, Crime, Drama, Fantasy, Romance, Sci-Fi.
\end{itemize}

\subsection{Success Cases: Personalized Attention Steering}
\label{sec:supp_success}

The first two rows of Figure~\ref{fig:supp_attention_examples} demonstrate sessions where \sysname successfully redirects the model's visual attention toward the user's characteristic gaze pattern.

\paragraph{KINIT\_49 / task 17 (Figures~\ref{fig:supp_sub_a}--\ref{fig:supp_sub_c}).}
The \backbone model (Figure~\ref{fig:supp_sub_a}) distributes attention relatively broadly across the grid, with moderate peaks in the upper-left region---consistent with a generic top-left primacy bias rather than any user-specific preference. In contrast, the human gaze map (Figure~\ref{fig:supp_sub_b}) is sharply concentrated on a small number of slots, with the dominant fixation falling on a single poster in the middle row. This user's profile indicates a strong preference for Sci-Fi and Fantasy, suggesting that the concentrated gaze reflects content-driven selective attention toward genre-relevant posters rather than a position-driven scanning habit. After personalized tuning, \sysname (Figure~\ref{fig:supp_sub_c}) substantially suppresses the diffuse background activation present in \backbone and shifts the peak relevance toward the same high-fixation region identified by the human gaze. The resulting attention distribution is noticeably more peaked and spatially aligned with the user's viewing pattern, illustrating the intended effect of personalized fixation alignment.

\paragraph{UvA\_25 / task 06 (Figures~\ref{fig:supp_sub_d}--\ref{fig:supp_sub_f}).}
This session presents a qualitatively different gaze pattern from the previous case. The \backbone attention (Figure~\ref{fig:supp_sub_d}) again exhibits a diffuse, position-biased distribution skewed toward the upper-left area. The human gaze (Figure~\ref{fig:supp_sub_e}), however, is distributed more broadly across several slots spanning both the upper and middle rows, with no single dominant peak---a pattern consistent with this user's wider range of preferred genres (six categories). Despite the more dispersed nature of this user's gaze, \sysname (Figure~\ref{fig:supp_sub_f}) successfully adjusts the model's attention: relevance is redistributed away from the generic upper-left concentration and toward the multiple slots that the user actually fixated on. This case is particularly informative because it demonstrates that the personalized prompt basis decomposition can capture not only sharp, single-peak gaze patterns but also broader, multi-focus browsing strategies---reflecting the diversity of individual viewing behaviors that our framework is designed to accommodate.

\paragraph{Cross-user contrast.}
Comparing the two success cases highlights the personalization effect. KINIT\_49 exhibits a narrow, content-driven gaze concentrated on genre-relevant items, while UvA\_25 displays a broader exploratory pattern spanning multiple candidates. The \backbone model produces similar diffuse attention maps for both users, failing to distinguish between these distinct browsing strategies. \sysname, by contrast, learns different personalized coefficients $\boldsymbol{\alpha}_u$ for each user, steering the shared prompt basis toward user-appropriate attention profiles. This cross-user divergence provides instance-level evidence that the factorized prompt design (Sec.~\ref{sec:prompt_basis} in the main paper) effectively disentangles shared attentional prototypes from individual viewing habits.

\subsection{Failure Case: Limitations of Attention Steering}
\label{sec:supp_failure}

The third row of Figure~\ref{fig:supp_attention_examples} presents a session where \sysname's alignment remains limited, offering insight into the current boundaries of the approach.

\paragraph{KINIT\_21 / task 15 (Figures~\ref{fig:supp_sub_g}--\ref{fig:supp_sub_i}).}
The human gaze (Figure~\ref{fig:supp_sub_h}) is sharply peaked on a single slot in the middle row, indicating a highly decisive viewing pattern. The \backbone (Figure~\ref{fig:supp_sub_g}) shows a different concentration that does not align with this peak. After tuning, \sysname (Figure~\ref{fig:supp_sub_i}) does shift attention partially toward the human-fixated region, but fails to fully suppress the competing high-relevance slots present in the \backbone---the model's top-ranked slot does not match the user's most-fixated slot, resulting in a CSH@1 miss.

We identify two plausible contributing factors. First, this user's profile lists seven preferred genres (the broadest among the three cases), which may make it harder for the personalized coefficient to converge toward a distinctive attentional mode given limited per-user training data (this user has relatively few training sessions). Second, the slot receiving the user's peak fixation may conflict with the VLM's content-level prior: if the model assigns low semantic relevance to that item based on the poster content, the soft prompt may lack sufficient capacity to override a strong content-driven bias in the frozen backbone. This suggests that attention steering through soft prompts alone may be insufficient when the target fixation pattern strongly contradicts the model's content-level representations---a limitation that could potentially be addressed by incorporating lightweight adapter layers or by jointly fine-tuning a small subset of backbone parameters in future work.

\subsection{Summary}

The extended qualitative analysis reveals three key insights. First, \sysname can accommodate diverse gaze patterns---from narrow single-focus viewing to broad multi-slot exploration---demonstrating the expressiveness of the personalized prompt basis decomposition. Second, the \backbone model produces largely undifferentiated attention maps across users, confirming the need for personalized alignment. Third, the failure case identifies a meaningful boundary condition: when the user's fixation target conflicts with the frozen model's content-level prior and when per-user training data is sparse, soft-prompt-based steering alone may be insufficient. This points toward promising future directions such as combining attention alignment with lightweight parameter adaptation.

%% file: Tables/supp_recgaze.tex
\begin{table*}[t]
\captionsetup{skip=1.3pt}
\caption{Extended evaluation on RecGaze with \mbox{standard deviations} and \mbox{significance levels} (cf.\ Table~1 in the main paper). $^{*}$\,$p<0.05$, $^{**}$\,$p<0.01$ (\mbox{two-sided} paired $t$-test, $\mathrm{df}=4$). Values are reported as mean\,$\pm$\,std.
\backbone values are deterministic (frozen model, greedy decoding) and thus reported without variance.
Bold indicates improvement over \backbone.}
\label{tab:supp_recgaze}
\centering
\scriptsize
\setlength\tabcolsep{3pt}
\renewcommand{\arraystretch}{0.95}
\resizebox{0.95\textwidth}{!}{%
\begin{tabular}{@{}l l l c c c c c c@{}}
\toprule
\multirow{2}{*}{\textbf{Model}} & \multirow{2}{*}{\textbf{Method}} & \multirow{2}{*}{\textbf{Setting}}
& \multicolumn{3}{c}{\textbf{Attention alignment metrics}}
& \multicolumn{3}{c}{\textbf{Prediction-level metrics}} \\
\cmidrule(lr){4-6}\cmidrule(lr){7-9}
& & & \textbf{CSH@1} & \textbf{TGO@1} & \textbf{KL}$\downarrow$ & \textbf{AUC} & \textbf{LogLoss}$\downarrow$ & \textbf{Accuracy} \\
\midrule
\multirow{6}{*}{\shortstack[l]{Qwen3-VL-\\4B-Instruct}}
& \multirow{2}{*}{GLIMPSE}
& \backbone & 0.1447 & 0.1447 & 1.1033 & 0.5893 & 2.7860 & 0.1053 \\
& & \sysname & \textbf{0.2368$\pm$0.053$^{*}$} & \textbf{0.1974$\pm$0.029$^{*}$} & \textbf{0.9986$\pm$0.060$^{*}$} & \textbf{0.6607$\pm$0.008$^{**}$} & \textbf{2.6042$\pm$0.082$^{**}$} & \textbf{0.1974$\pm$0.045$^{**}$} \\
\cmidrule(lr){2-9}
& \multirow{2}{*}{AttnLRP}
& \backbone & 0.0395 & 0.0658 & 1.2187 & 0.5893 & 2.7860 & 0.1053 \\
& & \sysname & \textbf{0.1316$\pm$0.038$^{**}$} & \textbf{0.1184$\pm$0.029$^{*}$} & \textbf{1.0954$\pm$0.056$^{**}$} & \textbf{0.6701$\pm$0.009$^{**}$} & \textbf{2.5759$\pm$0.081$^{**}$} & \textbf{0.2105$\pm$0.047$^{**}$} \\
\cmidrule(lr){2-9}
& \multirow{2}{*}{\shortstack[l]{Attention\\Rollout}}
& \backbone & 0.0921 & 0.0658 & 1.7114 & 0.5893 & 2.7860 & 0.1053 \\
& & \sysname & \textbf{0.1579$\pm$0.009$^{**}$} & \textbf{0.1711$\pm$0.038$^{**}$} & \textbf{1.1117$\pm$0.034$^{**}$} & \textbf{0.6513$\pm$0.017$^{**}$} & \textbf{2.6208$\pm$0.029$^{**}$} & \textbf{0.1711$\pm$0.019$^{**}$} \\
\midrule
\multirow{6}{*}{\shortstack[l]{InternVL3.5-\\4B-Instruct}}
& \multirow{2}{*}{GLIMPSE}
& \backbone & 0.0395 & 0.0526 & 1.5384 & 0.5203 & 2.7958 & 0.1184 \\
& & \sysname & \textbf{0.0789$\pm$0.019$^{**}$} & \textbf{0.1053$\pm$0.029$^{*}$} & \textbf{1.1978$\pm$0.086$^{**}$} & \textbf{0.5724$\pm$0.006$^{**}$} & \textbf{2.6864$\pm$0.046$^{**}$} & \textbf{0.1316$\pm$0.000$^{**}$} \\
\cmidrule(lr){2-9}
& \multirow{2}{*}{AttnLRP}
& \backbone & 0.0526 & 0.0395 & 1.7537 & 0.5203 & 2.7958 & 0.1184 \\
& & \sysname & \textbf{0.1053$\pm$0.026$^{*}$} & \textbf{0.1184$\pm$0.041$^{*}$} & \textbf{1.5118$\pm$0.053$^{**}$} & \textbf{0.6344$\pm$0.013$^{**}$} & \textbf{2.6088$\pm$0.084$^{**}$} & \textbf{0.1711$\pm$0.025$^{**}$} \\
\cmidrule(lr){2-9}
& \multirow{2}{*}{\shortstack[l]{Attention\\Rollout}}
& \backbone & 0.0395 & 0.0526 & 1.5791 & 0.5203 & 2.7958 & 0.1184 \\
& & \sysname & \textbf{0.1053$\pm$0.031$^{**}$} & \textbf{0.0658$\pm$0.000$^{**}$} & \textbf{1.2137$\pm$0.045$^{**}$} & \textbf{0.6136$\pm$0.011$^{**}$} & \textbf{2.6759$\pm$0.054$^{**}$} & \textbf{0.1974$\pm$0.034$^{**}$} \\
\bottomrule
\end{tabular}%
}
\end{table*}

%% file: Tables/supp_adserp.tex
\begin{table*}[t]
\captionsetup{skip=1.3pt}
\caption{Extended evaluation on AdSERP (AttnLRP) with \mbox{standard deviations} and \mbox{significance levels} (cf.\ Table~3 in the main paper). $^{*}$\,$p<0.05$, $^{**}$\,$p<0.01$ (\mbox{two-sided} paired $t$-test, $\mathrm{df}=4$).}
\label{tab:supp_adserp}
\centering
\scriptsize
\setlength\tabcolsep{3pt}
\renewcommand{\arraystretch}{0.95}
\resizebox{0.95\textwidth}{!}{%
\begin{tabular}{@{}l l c c c c c c@{}}
\toprule
\multirow{2}{*}{\textbf{Model}} & \multirow{2}{*}{\textbf{Setting}}
& \multicolumn{3}{c}{\textbf{Attention alignment metrics}}
& \multicolumn{3}{c}{\textbf{Prediction-level metrics}} \\
\cmidrule(lr){3-5}\cmidrule(lr){6-8}
& & \textbf{CSH@1} & \textbf{TGO@1} & \textbf{KL}$\downarrow$ & \textbf{AUC} & \textbf{LogLoss}$\downarrow$ & \textbf{Accuracy} \\
\midrule
\multirow{2}{*}{\shortstack[l]{Qwen3-VL-\\4B-Instruct}}
& \backbone & 0.2727 & 0.3939 & 0.5797 & 0.6760 & 1.4450 & 0.2727 \\
& \sysname & \textbf{0.3636$\pm$0.048$^{*}$} & \textbf{0.5758$\pm$0.057$^{**}$} & \textbf{0.5547$\pm$0.023} & \textbf{0.6851$\pm$0.006$^{*}$} & \textbf{1.3516$\pm$0.025$^{**}$} & \textbf{0.3636$\pm$0.030$^{**}$} \\
\midrule
\multirow{2}{*}{\shortstack[l]{InternVL3.5-\\4B-Instruct}}
& \backbone & 0.2121 & 0.4242 & 0.8482 & 0.5048 & 1.8929 & 0.2121 \\
& \sysname & \textbf{0.2727$\pm$0.043$^{*}$} & \textbf{0.4848$\pm$0.048$^{*}$} & \textbf{0.7355$\pm$0.035$^{**}$} & \textbf{0.5547$\pm$0.006$^{**}$} & \textbf{1.7947$\pm$0.040$^{**}$} & \textbf{0.2727$\pm$0.037$^{*}$} \\
\bottomrule
\end{tabular}%
}
\end{table*}